\documentclass{aa}
\usepackage{psfig}
\psfigurepath{Figures}
\begin{document}
\title{The AGN contribution to mid-infrared surveys}
\subtitle{X-ray counterparts of the mid-IR sources in the Lockman Hole and HDF
\thanks{Partially based on observations with ISO and XMM, ESA projects with instruments
funded by ESA Member States  with the participation of ISAS and NASA, and on observations
with the Canada--France--Hawaii Telescope at Mauna Kea, Hawaii.}}
\author{D. Fadda\inst{1,2} \and  H. Flores\inst{2,3} \and
        G. Hasinger\inst{4} \and  A. Franceschini\inst{5} \and B. Altieri\inst{6} \and 
        C.J. Cesarsky\inst{7} \and 
        D. Elbaz\inst{2} \and Ph. Ferrando\inst{2}      
        }
\offprints{D. Fadda}

\institute{Instituto de Astrof\'\i sica de Canarias (IAC), Via Lactea S/N, E-38205 La Laguna, Tenerife, Spain,\\ \email{fadda@ll.iac.es}
        \and
        CEA Saclay - Service d'Astrophysique, Orme des Merisiers, F91191 Gif-sur-Yvette Cedex, France
        \and
        Observatoire de Paris Meudon, DAEC, F92195 Meudon Principal Cedex, France
        \and
        AIP, An der Sternwarte 16, D14482 Potsdam, Germany
        \and
        Dipartimento di Astronomia, Universit\`a di Padova, Vicolo dell'Osservatorio 5, I35122 Padova, Italy
        \and    
        XMM-Newton Operations Centre, ESA Vilspa, Apartado 50727, E-28080 Madrid, Spain
        \and
        ESO, Karl-Schwarzschild Stra\ss e 2, D85748 Garching bei M\"unchen, Germany
             }

   \date{Received -; accepted -}

   \abstract{We provide constraints on the AGN contribution to the
        mid-IR extragalactic background light from a correlation
        analysis of deep X-ray and mid-IR observations in two regions
        centred on the Lockman Hole and Hubble Deep Field North (HDFN).  The
        Lockman region, of more than 200 square arcminutes, was observed
        by ISOCAM and XMM--Newton to a depth of 0.3 mJy at 15 $\mu$m
        (resolving more than 30\% of the mid-IR
        background).  In the same area XMM--Newton reached  flux
        limits of 1.4 $\times 10^{-15}$ erg cm$^{-2}$ s$^{-1}$ in the
        2--10 keV energy band and 2.4 $\times 10^{-15}$ erg cm$^{-2}$
        s$^{-1}$ in the 5--10 keV energy band, resolving about 80\%
        of the 2--10 keV  and 60\% of the 5--10 keV backgrounds
	(the deepest observation in this hard band to date).
        Among the 76 galaxies detected by XMM--Newton, 24 show mid-IR
        emission, but the relative percentage of X-ray sources with
        mid-IR counterparts increases with the band energy: from 30\%
        of the 0.5--2 keV sources up to 63\% of the 5--10 keV
        sources. In contrast, only a small fraction of the mid-IR
        sources (around 10\%) show X-ray emission within the
        sensitivity limits of XMM--Newton observations. The region
        centred on the HDFN has been observed by ISOCAM (24 square
        arcminutes) to a depth of 0.05 mJy (more than 50\% of the
        mid-IR background is resolved at this limit) and covered with
        a 1 Msec exposure by Chandra.  In this case, 25\% of the
        mid-IR sources are detected in the X-ray, while 30--40\% of the
        X-ray sources show mid-IR emission.  Under the conservative
        assumption that all XMM sources except stars or galaxy clusters
        are AGN--dominated, AGNs contribute $(15\pm5)\%$ of the total
        mid-IR flux in the Lockman Hole. For the HDFN we have assumed
        that AGN--dominated sources are luminous X-ray sources and
        sources with SEDs from radio to X-ray wavelengths typical of local 
        AGNs, in
        which case we find that $(18\pm7)\%$ of the mid-IR flux are
        due to AGN emission.  If we put together all the existing
        information from the deepest HDFN data to the bright
        large-area sample in the ELAIS S1 region observed with
        BeppoSAX (for a total of 50 X-ray--mid-IR matched sources) 
        using the median mid-IR to X-ray spectral indices as a
        function of the X-ray flux, we find an AGN contribution to the
        15 $\mu$m background of $(17\pm 2)\%$. This figure should be
        taken as an upper limit to the AGN contribution to the CIRB
        energy density, assuming standard SEDs for IR sources.  We
        conclude that the population of IR luminous galaxies detected
        in the ISOCAM deep surveys, and the CIRB sources themselves,
        are mainly constituted by dust-obscured starbursts.
        \keywords{cosmology --- infrared observations --- X-ray
        observations } } \maketitle

%

\section{Introduction}
The discovery of luminous mid-IR sources in the deep ISOCAM surveys
(Aussel et al. 1999; Elbaz et al. 1999) sheds new light on the star
formation history of the Universe (Rowan-Robinson et al. 1997; Flores
et al. 1999) and the origin of the infrared extragalactic background
light (Puget et al. 1998; Elbaz et al. 2001).  Mid-IR observations
appear to be fundamental to our understanding of 
 the evolution of the Universe because
they offer views of obscured star formation with a resolution
sufficient to identify optical counterparts of the infrared emitters,
which is often not the case of far-IR and sub-mm observations.  In
particular, the ISOCAM 15$\ \mu$m-centred band has proven to be well
suited to the study of star formation up to redshifts of 1.5 (Elbaz et
al. 2001), as well as to the resolution of a large part of the infrared
background detected by the DIRBE and FIRAS experiments (Puget et
al. 1998; Fixsen et al. 1998; Hauser et al. 1998).  Since a large part
of the star formation is detectable only in the infrared and ISOCAM
sources are rare with respect to optical/UV sources, this implies that
a large part of star forming activity occurs in rare, very luminous
systems.

Obscured AGNs are known to be the major
contributors to the hard X-ray background (e.g. Comastri et
al. 1995) and a large fraction of their optical--UV energy is
re-radiated at longer wavelengths, from the near-infrared to
the sub-millimetre region. Although the shape of the infrared spectrum of
such sources is unknown, some authors claim that absorbed AGNs could
contribute a substantial fraction (up to 50\%) of the infrared
background (Almaini et al. 1999; Fabian \& Iwasawa 1999).  However,
the nature of the energy source in powerful infrared emitters is still
a matter of debate. Recent ISO results suggest that most of these
galaxies could host an AGN, that but their emission is dominated by vigorous
star formation (Genzel et al. 1998; Lutz et al. 1998). Tran et
al. (2001), analysing almost 80 mid-IR spectra of luminous and
ultraluminous infrared galaxies, found that the average contributions
of star formation to the infrared luminosity are 82--94\% for 
low-luminosity sources ($L_{\rm IR}< 10^{12.4}\ L_{\odot}$) and 44--55\% for
high luminosity sources ($L_{\rm IR}\ge 10^{12.4}\ L_{\odot}$).

If the bulk of the ISOCAM 15$\ \mu$m galaxies, which have a typical
luminosity in the $10^{11}\ L_{\odot}<L_{\rm IR}< 10^{12}\ L_{\odot}$ range
and make up most of the infrared background (Elbaz et al. 2001), were
dusty starbursts, this would have important implications on the star
formation history of the Universe (see Flores et al. 1999).

When detailed optical spectra are not available, it is in principle
possible to classify mid-IR galaxies using mid-IR based diagnostic
diagrams (Laurent et al. 2000) or fitting templates with
multi-wavelength data (Flores et al. 1999).  The major drawback of the
multi-wavelength approach is the similarity of starburst and Seyfert-2
SEDs in the radio--optical domain (see examples in Flores et al.
1999). On the other hand, the mid-IR diagnostic by Laurent et
al. (2000), which is based on only five local AGNs, strongly depends
on the source redshift. Moreover, in the case of deep surveys, large
photometric errors make it difficult to distinguish between starburst-
and AGN-dominated mid-IR sources.

A direct approach consists in cross-correlating mid-IR and X-ray observations
to detect AGNs and, in particular, using hard X-ray energy bands which are less
sensitive to the extinction and are thus able to reveal obscured AGNs.

In this perspective, we compared a recent XMM--Newton observation (Hasinger
et al. 2001) with the mid-IR ISOCAM surveys at 6.75$\ \mu$m and 15$\ \mu$m (Fadda
et al. 2002) of a region located in the Lockman Hole and the deep 
Chandra observation (Brandt et al. 2001a) with the ISOCAM survey (Aussel et 
al. 1999) of a region centred on the Hubble Deep Field.

Pioneering studies of this type have been already tried by comparing
BeppoSAX with ISOCAM data in the Elais-S1 field (Alexander et al. 2001)
and Chandra with ISOCAM data in the HDFN field (Hornschemeier et al. 2001).
However, these studies are respectively affected  by the low sensitivity of the
surveys and  the small area observed.
The present one is the first study with a sufficient depth and sky coverage to
obtain a statistically significant set of sources with hard X-ray and
mid-IR emission.

In this paper, after discussing the
cross-correlation of mid-IR and X-ray catalogues and the common optical,
mid-IR and X-ray properties of the sources,
 we compare the sources detected in the surveys made in the
Lockman, HDFN and Elais-S1 regions with ISOCAM and different
X-ray satellites (XMM, Chandra and Beppo-SAX) with local templates.
Finally, we estimate the AGN contribution to the extragalactic
infrared light.


\section{Observations}

\begin{figure*}[!t]
\hbox{
\vspace*{9cm}
}
\caption{Relative positions of the mid-IR and X-ray surveys.  On the
left, on the full-band XMM image of the Lockman Hole we superimpose
the contours of the area deeply surveyed by ISOCAM and the circle
from inside  which Hasinger et al. (2001) sources have been extracted.
The area considered in the paper results from the intersection of
these two regions.  On the right, on the full-band Chandra image of
the Hubble Deep Field and flanking fields we draw: the field observed
with the Hubble Space Telescope, the field surveyed with ISOCAM (irregular
contour) and the Caltech area (square).  }%
\label{fig:surveys}%
\end{figure*}

A region of $20 \times 20$ square arcminutes in the Lockman Hole,
centred on the sky position 10:52:07+57:21:02 (2000), which
corresponds to the centre of the ROSAT HRI image (Hasinger et
al. 1993, 1998), has been surveyed by ISOCAM, the mid-IR camera on board ISO
(Fadda et al. 2002). The field was observed for a total of 45 ks
at 15$\ \mu$m and 70 ks at 6.75$\ \mu$m. Moreover, a shallow survey
has been done at 15$\ \mu$m on a region of $40 \times 40$ arcmin$^2$ with
the same centre for a total exposure time of 55 ks. Combining the
two surveys, ISOCAM observed at 15$\ \mu$m the central region for a
total of 60 ks.

If we compare this observation with the deepest ISOCAM surveys
performed in the HDFN region (24.3 square arcminutes for 22 ks
at 15$\ \mu$m and 10.4 square arcminutes for 23 ks at 6.75$\ \mu$m,
Aussel et al. 1999), the Lockman Deep Survey is 16 times more extended
and $\sim$ 4 times shallower at 15$\ \mu$m than the HDF survey (the
sensitivity depending on the integration time and on the redundancy of
the observations).  ISOCAM data have been reduced using the PRETI
pipeline (see Starck et al. 1999) with a few improvements as described
by Fadda et al. (2000).

The  XMM--Newton observation of the Lockman Hole has been done during
the verification phase of the satellite (Hasinger et al. 2001) and is 
centred on the sky position 10:52:43+57:28:48 (2000) which was the
centre of the PSPC ROSAT image (Hasinger et al. 1998). The total
exposure time of this observation was 190 ks, but only 100 ks
are usable because of bad space weather (solar activity) during the
observations. The limiting fluxes of these observations are
$3\times 10^{-16}$ erg cm$^{-2}$ s$^{-1}$ in the 0.5--2 keV band
and $1.4\times 10^{-15}$ erg cm$^{-2}$ s$^{-1}$ in the 2--10 keV band.

To match X-ray and mid-IR sources we considered X-ray sources detected
at the 4$\sigma$ level inside an off-axis angle of 10 arcminutes (Hasinger
et al. 2001).

Due to the different centre of the ISOCAM and XMM--Newton observations,
the size of the overlapping region is 218 square arcminutes (see
Fig.~\ref{fig:surveys}), which corresponds to 70\% of the XMM--Newton
region.  In this region, a total of 76 sources was detected
in the various X-ray bands excluding clusters and stars (68, 42 and
19 in the 0.5--2 keV, 2--10 keV and 5--10 keV energy bands,
respectively).  In the same area,  184 and 65 extragalactic sources were detected by
ISOCAM at the 3$\sigma$ level in the LW3 band and at 4$\sigma$ level in the LW2 band which
are centred on 15$\ \mu$m and  6.75$\ \mu$m, respectively (Fadda et al. 2002). 
\begin{table}[b!]
\setlength{\tabcolsep}{2.9mm}
\begin{tabular}{ccccccc}
\hline
XMM &  \multicolumn{3}{c}{LW3} & \multicolumn{3}{c}{LW2} \\
band &  \# & X\%  & IR\% &  \# & X\%  & IR\% \\ 
\hline
Soft  & 20 & 29$^{+12}_{-10}$ & 11$^{+3}_{-3}$ & 6 &  9$^{+6}_{-5}$   &  9$^{+7}_{-5}$ \\
Hard  & 16 & 38$^{+19}_{-15}$ &  9$^{+6}_{-4}$ & 5 & 12$^{+10}_{-7}$  &  8$^{+6}_{-4}$ \\
U-hard& 12 & 63$^{+42}_{-32}$ &  7$^{+3}_{-2}$ & 5 & 26$^{+25}_{-17}$ &  8$^{+6}_{-4}$ \\
Full  & 22 & 29$^{+11}_{-9}$  & 12$^{+7}_{-5}$ & 7 &  9$^{+6}_{-4}$   & 11$^{+7}_{-5}$ \\
\hline
\end{tabular}
\caption{Number of X-ray sources detected at 15$\ \mu$m and 6.75$\
\mu$m in the Lockman Hole-centred region and percentage with
one-sigma errors of common detections relative to the total number of
X-ray and mid-IR sources in the different XMM--Newton bands. Poissonian
errors are computed according to Gehrels et al. (1986).}
\label{tab:stat_lock}
\end{table}
As shown in Fig.~\ref{fig:surveys}, the region centred on the
Hubble Deep Field observed by ISOCAM (Rowan-Robinson et al. 1997) has
been completely covered by deep 1~Ms Chandra observations (Brandt et
al. 2001a).  Chandra observations are one order of magnitude deeper
than the XMM--Newton observations in the 0.5--2 keV band (flux limit of
$3\times 10^{-17}$ erg cm$^{-2}$ s$^{-1}$) and 2--8 keV band (flux
limit of $2\times 10^{-16}$ erg cm$^{-2}$ s$^{-1}$).  On the contrary,
Chandra is less sensitive than XMM--Newton in the ultra-hard band ($>
5$ keV).  In the X-ray--LW3 common area, Chandra detects a total of 59 sources in the
full X-ray band (0.5--8 keV) and 50, 40 and 23 sources in the 0.5--2
keV, 2--8 keV and 4-8 keV bands, respectively.  Aussel et al. (1999)
list a total of 93 LW3 sources, 42 of which have a flux greater than
0.1 mJy (completeness flux limit of the survey, Aussel et al. 2001).
In the X-ray--LW2 common area (10.4 square arcminutes with 10 ISOCAM detections),
Chandra detects a total of 24 sources in the full band and 22, 12 and 7 sources
in the soft, hard and ultra-hard bands, respectively.

\begin{table}[b!]
\setlength{\tabcolsep}{2.7mm}
\begin{tabular}{ccccccc}
\hline
Chandra &  \multicolumn{3}{c}{LW3} & \multicolumn{3}{c}{LW2} \\
band &  \# & X\%  & IR\% &  \# & X\%  & IR\% \\ 
\hline
Soft  & 20 & 40$^{+18}_{-15}$ & 21$^{+8}_{-7}$& 5&  23$^{+21}_{-15}$&50$^{+55}_{-37}$ \\
Hard  & 12 & 30$^{+17}_{-13}$ & 13$^{+6}_{-5}$& 2&  17$^{+10}_{-6}$ &20$^{+35}_{-19}$ \\
U-hard&  7 & 30$^{+24}_{-18}$ &  8$^{+5}_{-4}$& 2&  29$^{+6}_{-4}$  &20$^{+35}_{-19}$ \\
Full  & 22 & 37$^{+15}_{-13}$ & 24$^{+9}_{-7}$& 5&  21$^{+19}_{-13}$&50$^{+55}_{-37}$ \\
\hline
\end{tabular}
\caption{Number of X-ray sources detected at 15$\ \mu$m and 6.75$\
\mu$m in the Hubble Deep Field-centred region and percentage with
one-sigma errors of common detections relative to the total number of
X-ray and mid-IR sources in the different Chandra bands. Poissonian
errors are computed according to Gehrels et al. (1986).}
\label{tab:stat_hdf}
\end{table}

\begin{table*}[ht!]
\setlength{\tabcolsep}{0.3mm}
\begin{tabular}{lcc cccccc cccc rrrrr ccc rcc}
\hline
\multicolumn{1}{c}{Names}&
\multicolumn{2}{c}{J2000 Coords}&
\multicolumn{2}{c}{X-Op}&
\multicolumn{2}{c}{IR-Op}&
\multicolumn{2}{c}{X-IR}&
\multicolumn{4}{c}{Optical}&
\multicolumn{6}{c}{X-ray}&
\multicolumn{2}{c}{mid-IR}&
&&\\
&
\multicolumn{1}{c}{$\alpha$-10$^h$}&
\multicolumn{1}{c}{$\delta$-57$^o$}&
\multicolumn{2}{c}{$\Delta \ \ $ P}&
\multicolumn{2}{c}{$\Delta \ \ $ P}&
\multicolumn{2}{c}{$\Delta \ \ $ P}&
\multicolumn{1}{c}{V}&
\multicolumn{1}{c}{V-I}&
\multicolumn{1}{c}{R}&
\multicolumn{1}{c}{R-K'}&
\multicolumn{1}{c}{SX}&
\multicolumn{1}{c}{HX}&
\multicolumn{3}{c}{Hard. Ratios}&
\multicolumn{1}{c}{L$_X$}&
\multicolumn{1}{c}{LW2}&
\multicolumn{1}{c}{LW3}&
\multicolumn{1}{c}{$\alpha_{IX}$}&
\multicolumn{1}{c}{z}&
\multicolumn{1}{c}{T}\\
\multicolumn{1}{c}{(1)}&
\multicolumn{1}{c}{(2)}&
\multicolumn{1}{c}{(3)}&
\multicolumn{2}{c}{(4) (5)}&
\multicolumn{2}{c}{(6) (7)}&
\multicolumn{2}{c}{(8) (9)}&
\multicolumn{1}{c}{(10)}&
\multicolumn{1}{c}{(11)}&
\multicolumn{1}{c}{(12)}&
\multicolumn{1}{c}{(13)}&
\multicolumn{1}{c}{(14)}&
\multicolumn{1}{c}{(15)}&
\multicolumn{1}{c}{(16)}&
\multicolumn{1}{c}{(17)}&
\multicolumn{1}{c}{(18)}&
\multicolumn{1}{c}{(19)}&
\multicolumn{1}{c}{(20)}&
\multicolumn{1}{c}{(21)}&
\multicolumn{1}{c}{(22)}&
\multicolumn{1}{c}{(23)}&
\multicolumn{1}{c}{(24)}\\
\hline
$\ $1$\ $(32A)&52:39.6&24:32&0.6 &  0&0.3 &  0&0.1&0&17.8&0.6&17.5&1.4&58.7&*60.2&  .10& -.71& -.46&$\ $44.61& 0.9$^{+0.3}_{-0.3}$&1.3$^{+0.4}_{-0.3}$&   1.17& 1.113& 1\\
$\ $7$\ $(37A)&52:48.3&21:18&1.5 &  2&2.5 &  5&2.9&2&20.3&0.9&19.9&2.6&11.2&* 7.7& -.08& -.79& -.44&$\ $42.80& 0.3$^{+0.3}_{-0.3}$&0.6$^{+0.3}_{-0.3}$&   1.29& 0.467& 1\\
12$\ $(30A)   &52:57.2&25:07&1.4 &  3&2.6 & 12&2.7&2&21.5&1.2&20.9&2.5& 6.1&  7.2&  .16& -.69& -.62&$\ $44.06&              $<$0.3&0.6$^{+0.3}_{-0.4}$&   1.29& 1.527& 1\\
13(120A)      &53:09.5&28:21&1.0 &  1&1.7 &  3&2.1&1&20.7&0.9&20.4&2.0& 5.4&* 7.5&  .20& -.63& -.56&$\ $44.08&              $<$0.3&0.8$^{+0.4}_{-0.2}$&   1.32& 1.568& 1\\
14$\ $(45Z)   &53:19.1&18:53&0.5 &  0&3.6 & 15&4.0&5&21.3&2.3&21.2&-  & 4.7&  5.2&  .44& -.66& -.46&$\ $43.07& $<$0.3            &0.6$^{+0.3}_{-0.4}$&   1.32& 0.711& 2\\
15$\ $(14Z)   &52:42.6&31:58&2.0 & 96&2.1 &105&1.0&2&25.2&1.4&25.0&5.6& 4.6&*12.8&  .96& -.33& -.50&$\ $44.51&$<$0.3             &0.7$^{+.35}_{-0.2}$&   1.26& (1.9)& -\\ 
16$\ $(12A)   &51:49.6&32:49&2.5 & 41&2.8 & 51&0.9&0&23.2&-  &22.9&4.9& 4.1&*32.9&  .93&  .09& -.40&$\ $44.22& 0.3$^{+0.3}_{-0.2}$&0.4$^{+0.3}_{-0.3}$&   1.14& 0.990& 2\\
21(814A)      &52:45.3&21:23&3.8 & 23&1.0 &  2&4.0&8&20.9&0.7&20.4&1.5& 3.0&  2.9&  .19& -.66& -.67&$\ $44.29& $<$0.3            &0.3$^{+0.2}_{-0.2}$&   1.32& 2.832& 1\\
32(486A)      &52:43.4&27:59&1.0 & 10&1.6 & 25&0.8&0&24.7&2.2&24.4&5.6& 1.9&* 4.4&  .92& -.41& -.39&$\ $43.36&$<$0.3             &0.3$^{+0.3}_{-0.2}$&   1.31& (1.0)& -\\
35            &51:46.6&30:35& -  &  -& -  & - &4.0&8&  - & - &25.6&4.0& 1.8&-    &  .64& -.79& -.00&     -   &$<$0.3             &0.3$^{+0.2}_{-0.2}$&$>$1.40&  -   & -\\
36$\ $(36F)   &52:24.9&22:49&3.2 & 18&2.3 &  9&3.0&2&21.7&1.4&22.4&4.4& 1.8&  -  &  .22& -.56& -.82& $<$42.63&$<$0.3             &0.7$^{+0.3}_{-0.2}$&$>$1.45& 0.807& 1\\
41            &51:28.3&27:40&0.4 &  1&3.7 &120&4.0&8&24.3&1.9&22.8&3.2& 1.6&*12.4&  1.0&  .14& -.54&$\ $44.17& $<$0.3            &0.3$^{+0.3}_{-0.2}$&   1.22&(1.4) & -\\
43            &52:06.8&29:26&1.1 &  2&0.7 &  1&1.9&1&22.6&2.3&21.9&4.5& 1.5&* 8.4&  .87& -.01& -.68&$\ $43.33&$<$0.3             &0.6$^{+0.3}_{-0.3}$&   1.28&(0.75)& -\\
78(438A)      &52:55.1&19:52&3.0 & 59&0.4 &  1&3.3&0&24.6&2.6&24.2&6.0& 0.7&  -  &  .67& -.36& -.70& $<$43.23& 0.4$^{+0.3}_{-0.3}$&1.4$^{+0.4}_{-0.3}$&$>$1.52& (1.4)& -\\
79$\ $(33A)   &52:00.6&24:21&2.1 & 15&3.7 & 50&4.0&6&22.4&1.3&22.1&2.5& 0.7&* 2.0&  .77& -.32& -.08&$\ $42.99&$<$0.3             &0.5$^{+0.3}_{-0.4}$&   1.41& 0.974& 1\\
82            &51:55.0&24:07&2.0 & 12&0.4 &  0&1.9&0&22.7&1.7&22.0&3.3& 0.6&  -  &  .87& -.27& -.74& $<$43.80&              $<$0.3&1.0$^{+0.4}_{-0.2}$&$>$1.49& 2.4  & -\\
95(901A)      &52:52.7&29:00&1.2 &  0&0.2 &  0&1.3&0&17.8&1.3&18.1&3.9& 0.5&* 3.3& -.20&  .07&  .37&$\ $41.60& 0.7$^{+0.3}_{-0.3}$&2.4$^{+0.5}_{-0.5}$&   1.50& 0.204& 2\\
98            &51:36.9&29:45&3.0 & 55&1.4 & 12&4.0&8&23.0&1.1&22.8&-  & 0.5&  -  &  1.0& -.38& -.77&     -   &   $<$0.3          &0.3$^{+0.3}_{-0.2}$&$>$1.41& -    & -\\
104           &53:04.9&30:53&0.1 &  0&3.0 &  5&3.0&-&22.2&3.2&21.7&5.9& 0.4&  -  &  .36& -.96&  .84& $<$42.86& 0.3$^{+0.3}_{-0.2}$&$<0.7$                 &      -& (1.0)& -\\
105           &53:15.8&24:50&0.4 &  2&2.6 & 65&3.2&5&25.8&3.3&24.5&5.1& 0.4&  5.3&  .69&  .26& -.28&$\ $43.20&$<$0.3          &0.3$^{+0.3}_{-0.2}$&   1.30& (0.8)& -\\
109           &51:44.6&26:51&2.2 & 50&2.1 & 46&1.6&1&24.3&1.7&23.9&4.7& 0.4&  -  &  .90& -.64& -.82& $<$42.97&$<$0.3          &0.3$^{+0.2}_{-0.2}$&$>$1.40& (1.1)& -\\
113           &53:05.6&28:10&1.5 & 12&1.5 & 12&1.3&0&23.3&1.6&23.0&4.3&  - &* 8.8&   - &  .63& -.13&$\ $43.55&$<$0.3          &0.7$^{+0.4}_{-0.2}$&   1.30& (0.9)& -\\
115           &52:32.1&24:30&1.2 & 36&3.3 &140&3.1&5&26.4&2.6&24.7&5.3&  - &* 4.9&   - &  .35& -.18&$\ $43.51&$<$0.3          &0.3$^{+0.3}_{-0.2}$&   1.29& (1.1)& -\\
121           &52:31.4&25:04&3.5 &195&3.4 &205&0.5&-&25.3&1.9&24.9&5.1&  - &* 2.3&   - &  .51&  .07&$\ $43.08& 0.3$^{+0.3}_{-0.2}$&$<0.5$                 &      -& (1.0)& -\\
\hline
\end{tabular}
\caption{List of the XMM-Newton sources with mid-IR counterparts in
the Lockman Hole region ordered with decreasing 0.5--2 keV flux.  For
each association we report: the XMM number with previous ROSAT name
within brackets (1); the coordinates of the X-ray source (2,3); the
offset between associated sources in arcseconds and the probability of
random association in $10^{-3}$ units for X-ray vs. optical (4,5),
mid-IR vs. optical (6,7) and X-ray vs. mid-IR (8,9), respectively; the V, 
I, R and K$'$ magnitudes based on CFHT and Calar Alto observations
(10--13); the X-ray fluxes in the soft (0.5--2 keV) and hard (2--10 keV)
X-ray bands in $10^{-15}$ cgs units (14,15); the hardness ratios
(16--18); the 2--10 keV luminosity (19); the mid-IR fluxes at 6.75$\
\mu$m (20) and 15$\ \mu$m in mJy (21); the $\alpha_{\rm IX}$ index (22)
described in the text; the measured redshift (23) and the AGN type
(24). Zero probability of random association means less than $0.5
\times 10^{-3}$.  The hardness ratios, defined as $(H-S)/(H+S)$ (where
$H$ and $S$ represent hard and soft bands, respectively), compare 0.2--0.5
vs. 0.5--2 keV, 0.5--2 vs. 2--4.5 keV and 2--4.5 vs. 4.5--10 keV,
respectively.  In three cases a 15$\ \mu$m source has been marginally
detected (SNR $<$ 4) at 6.75$\ \mu$m.  We report their 6.75$\ \mu$m
fluxes inside brackets. Detections in the 5--10 keV X-ray bands are
marked with an asterisk on the 2--10 keV flux. The 6.75$\ \mu$m flux of
the source \# 95 has been derived by deblending two close
sources. Redshifts are taken from Lehmann et al. (2000, 2001) except
for \# 82 (Fadda et al. 2002). Photometric redshifts computed on the
basis of V, I, R and K$'$ magnitudes are reported inside brackets.}
\label{tab:list}
\end{table*}

Tables~\ref{tab:stat_lock} and~\ref{tab:stat_hdf} summarise the
percentages of X-ray and mid-IR sources which emit in the mid-IR and
X-ray bands, respectively.  It appears clear that a large fraction of
X-ray sources have an LW3 counterpart.  In particular, in the case of
the Lockman Hole, the percentage of X-ray sources emitting in the 5--10
keV band with LW3 counterpart is greater than 60\%.  The same does not
occur in the case of the Chandra deep field, probably because the 4--8
keV ultra-hard band of Chandra is not so sensitive as the similar band
of XMM-Newton.  On the other hand, only around 10\% of the LW3 sources
are detected in the various X-ray bands except for the soft X-ray band
in the HDF, where the extremely deep Chandra observations are able also to
detect  normal galaxies.

In the case of LW2 observations, we have less detections with respect to LW3
and a similar trend of detections as a function of the energy band in the 
Lockman Hole. The case of the Hubble Deep Field is not very interesting 
because of  the  bad quality of the LW2 observations (only 10 
extragalactic sources have been detected).

\section{Analysis}

\begin{table*}[ht!]
\setlength{\tabcolsep}{0.2mm}
\begin{tabular}{l cc cccccc cccc rrr cc rccc}
\hline
\multicolumn{1}{c}{Names}&
\multicolumn{2}{c}{J2000 Coords}&
\multicolumn{2}{c}{X-Op}&
\multicolumn{2}{c}{IR-Op}&
\multicolumn{2}{c}{X-IR}&
\multicolumn{4}{c}{Optical}&
\multicolumn{3}{c}{X-ray}&
\multicolumn{2}{c}{mid-IR}&
&&&\\
&
\multicolumn{1}{c}{$\alpha$-12$^h$}&
\multicolumn{1}{c}{$\delta$-62$^o$}&
\multicolumn{2}{c}{$\Delta \ \ $ P}&
\multicolumn{2}{c}{$\Delta \ \ $ P}&
\multicolumn{2}{c}{$\Delta \ \ $ P}&
\multicolumn{1}{c}{V}&
\multicolumn{1}{c}{V-I}&
\multicolumn{1}{c}{R}&
\multicolumn{1}{c}{R-Ks}&
\multicolumn{1}{c}{SX}&
\multicolumn{1}{c}{HX}&
\multicolumn{1}{c}{L$_X$}&
\multicolumn{1}{c}{LW2}&
\multicolumn{1}{c}{LW3}&
\multicolumn{1}{c}{$\alpha_{IX}$}&
\multicolumn{1}{c}{z}&
\multicolumn{1}{c}{C}&
\multicolumn{1}{c}{T}\\
\multicolumn{1}{c}{(1)}&
\multicolumn{1}{c}{(2)}&
\multicolumn{1}{c}{(3)}&
\multicolumn{2}{c}{(4) (5)}&
\multicolumn{2}{c}{(6) (7)}&
\multicolumn{2}{c}{(8) (9)}&
\multicolumn{1}{c}{(10)}&
\multicolumn{1}{c}{(11)}&
\multicolumn{1}{c}{(12)}&
\multicolumn{1}{c}{(13)}&
\multicolumn{1}{c}{(14)}&
\multicolumn{1}{c}{(15)}&
\multicolumn{1}{c}{(16)}&
\multicolumn{1}{c}{(17)}&
\multicolumn{1}{c}{(18)}&
\multicolumn{1}{c}{(19)}&
\multicolumn{1}{c}{(20)}&
\multicolumn{1}{c}{(21)}&
\multicolumn{1}{c}{(22)}\\
\hline
144 PM3\_6  &36:36.64&13:46.9&0.4& 0&1.5& 5&1.9&1&22.1&1.3& 20.8&   2.7&   4.43&*  5.53&$\ $44.02& -                    &0.35$^{+0.04}_{-0.07}$&$\ $1.26&0.957$^0$& $\mathcal{Q}$& 1\\
171 PM3\_20 &36:46.35&14:04.8&0.3& 0&1.2& 3&1.4&3&22.9&2.1& 21.7&   3.9&   2.80& *20.10&$\ $44.19&0.19$^{+0.04}_{-0.09}$&0.11$^{+0.09}_{-0.02}$&$\ $1.03&0.961$^2$& -            & 1\\
163 PS3\_10 &36:42.22&15:45.8&0.2& 0&1.1& 3&1.2&0&23.3&2.4& 21.6&   4.3&   0.84&*  2.48&$\ $43.32& -                    &0.46$^{+0.05}_{-0.09}$&$\ $1.35&0.857$^5$& $\mathcal{I}$& - \\
198 PS3\_24 &36:55.46&13:11.4&0.4& 1&0.9& 4&0.6&1&24.5&2.8& 22.9&   4.7&   0.40&*  0.89&$\ $43.53& $<$0.4               &0.02$^{+0.01}_{-0.01}$&$\ $1.16&1.315$^3$& $\mathcal{A}$   & - \\
190 PM3\_29 &36:51.75&12:21.4&1.3& 5&1.9&10&1.4&5&22.6&1.6& 21.5&   2.9&   0.28& * 2.59&$\ $42.27&$<$0.04               &0.05$^{+0.03}_{-0.01}$&$\ $1.14&0.401$^0$&$\mathcal{E}$& - \\
142 PM3\_5  &36:35.60&14:24.4&0.5& 2&0.3& 1&0.5&0&23.9&1.2& 23.5&   4.5&   0.28&*  2.82&$\ $44.11& -                    &0.44$^{+0.04}_{-0.08}$&$\ $1.33&2.011$^6$& -            & 2\\
134 PM3\_2  &36:34.46&12:12.9&0.3& 0&0.6& 0&1.0&0&21.1&2.0& 18.8&   2.7&   0.23&$<$0.27& $<$41.63&  -                   &0.45$^{+0.07}_{-0.06}$& $>$1.56&0.456$^0$& $\mathcal{I}$& -\\
176 PS2\_3  &36:48.05&13:09.1&0.3& 0&0.9& 1&0.9&2&22.5&2.5& 20.4&   3.2&   0.18&*  0.63&$\ $41.95&0.04$^{+0.06}_{-0.03}$&          $<$ 0.07    & $<$1.30&0.475$^2$& $\mathcal{I}$& - \\
172 PM3\_21 &36:46.41&15:29.2&0.2& 0&0.6& 2&0.7&0&24.6&3.1& 23.2&   4.2&   0.14&*  0.40&$\ $42.07& -                    &0.42$^{+0.09}_{-0.09}$&$\ $1.51&(0.6)    & -            & - \\
160 PM3\_12 &36:41.80&11:32.0&1.5& 2&1.4& 1&2.7&5&20.6&0.9& 19.4&   1.7&   0.13&$<$0.26& $<$39.78&        $<$0.07       &0.24$^{+0.06}_{-0.06}$& $>$1.50&0.089$^1$&  $\mathcal{EI}$ & -\\
136 PM3\_3  &36:34.51&12:41.6&0.7& 4&1.2&11&1.1&0&24.2&2.0& 23.2&   4.5&   0.10&$<$0.27& $<$42.87&  -                   &0.36$^{+0.08}_{-0.04}$& $>$1.53&1.219$^0$& $\mathcal{E}$& -\\
161 PS3\_6e &36:42.11&13:31.6& - & -& - & -&1.3&6&  - & - &$>$26&$>$4.6&   0.10&$<$0.21& $<$44.66&$<$0.07               &0.02$^{+0.01}_{-0.01}$& $>$1.29&4.424$^4$& $\mathcal{E}$& 1\\
188 PM3\_28 &36:51.11&10:30.7&0.3& 0&0.4& 0&0.5&0&21.9&2.0& 20.2&   3.6&   0.10&$<$0.26& $<$41.42&  -                   &0.34$^{+0.04}_{-0.07}$& $>$1.53&0.410$^1$& $\mathcal{I}$& -\\
194 PM3\_32 &36:53.41&11:39.6&0.6& 2&2.1&27&1.8&3&23.3&1.4& 23.2&   4.0&   0.09&$<$0.15& $<$42.67&  -                   &0.18$^{+0.06}_{-0.04}$& $>$1.52&1.275$^0$& $\mathcal{EA}$&-\\
Var PM3\_17 &36:44.20&12:51.0&1.9& 6&1.8& 6&0.2&0&22.6&2.1& 21.4&   2.1&   0.08&$<$0.29& $<$41.93&$<$0.05               &0.28$^{+0.06}_{-0.06}$& $>$1.50&0.557$^0$& $\mathcal{E}$& 2\\
155 PM3\_11 &36:40.00&12:50.2&0.8& 2&1.1& 4&0.9&1&23.2&2.2& 21.5&   3.6&   0.07&   0.27&$\ $42.37&$<$0.06               &0.30$^{+0.07}_{-0.06}$&$\ $1.52&0.848$^0$& $\mathcal{I}$& -\\
175 PS3\_14 &36:47.95&10:19.9& - & -& - & -&2.3&8&  - & - &  - &      -&   0.07&   0.89&     -   &  -                   &0.10$^{+0.09}_{-0.02}$&    1.30&  -      & -   & - \\
185 PM3\_27 &36:49.76&13:13.0&2.0&11&0.4& 0&2.0&6&23.0&2.0& 21.5&   3.4&   0.07&$<$0.17& $<$41.41&0.14$^{+0.07}_{-0.06}$&0.16$^{+0.04}_{-0.04}$& $>$1.51&0.475$^1$& $\mathcal{I}$ & - \\
178 PM3\_24 &36:48.38&14:26.2&0.7& 0&1.4& 1&1.1&1&19.5&1.0& 18.7&   2.0&   0.06&$<$0.17& $<$40.04&0.25$^{+0.07}_{-0.07}$&0.31$^{+0.06}_{-0.07}$& $>$1.56&0.139$^0$& $\mathcal{E}$& - \\
220 PM3\_42 &37:02.04&11:22.4&0.3& 0&2.0& 2&2.0&4&20.3&1.2& 18.7&   2.2&   0.06&$<$0.18& $<$40.04&  -                   &0.16$^{+0.08}_{-0.05}$& $>$1.50&0.136$^2$& $\mathcal{I}$& - \\
183 PM2\_3  &36:49.45&13:47.2&0.6& 0&0.8& 0&1.0&3&19.2&1.5& 18.0&   2.4&   0.05&$<$0.18& $<$39.62&0.04$^{+0.07}_{-0.03}$&          $<$ 0.05    &      - &0.089$^1$& -   & - \\
148 PM3\_7  &36:37.01&11:34.9&1.3& 1&1.4& 1&1.2&1&19.6&1.0& 17.7&   1.8&   0.04&$<$0.09& $<$39.19&$<$0.14               &0.30$^{+0.06}_{-0.07}$& $>$1.62&0.078$^1$& $\mathcal{I}$& -\\
229 PS3\_37 &37:04.66&14:29.0&0.2& 0&1.4& 3&1.5&5&22.7&2.3& 21.0&   3.2&$<$0.05&   0.29&$\ $41.86& -                    &0.07$^{+0.06}_{-0.02}$&$\ $1.38&0.561$^1$& $\mathcal{I}$& - \\
149 PS3\_6b &36:38.50&13:39.5&0.7& 2&1.0& 5&1.2&4&23.6&2.0& 22.3&   3.3&$<$0.04&   0.75&$\ $41.57& -                    &0.05$^{+0.03}_{-0.01}$&$\ $1.26&0.357$^5$& $\mathcal{I}$& - \\
\hline                                 
\end{tabular}                            
\caption{List of the Chandra sources with mid-IR counterparts in the
Hubble Deep Field and flanking fields ordered with decreasing 0.5--2
keV flux.  For each association we report: the number of Brandt et
al. (2001a) Chandra source and the name of the ISOCAM counterpart
(Aussel et al., 1999) (1); the J2000 coordinates (2,3) of the Chandra
source; the offset between associated sources in arcseconds and the
probability of random association in $10^{-3}$ units for X-ray vs.
optical (4,5), mid-IR vs. optical (6,7) and X-ray vs. mid-IR (8,9),
respectively; the V, I magnitudes (10,11) from Barger et al. (1999),
the R and Ks magnitudes (12,13) from Hogg et al. (2000); the X-ray
fluxes in the soft (0.5--2 keV) and hard (2--8 keV) X-ray bands in
10$^{-15}$ cgs units (14,15); the 2--10 keV luminosity (16); the mid-IR
fluxes at 6.75$\ \mu$m (17) and 15$\ \mu$m (18) in mJy (from Aussel et
al. 1999 and 2001); the $\alpha_{IX}$ index described in the text
(19); the measured redshift (20); the spectral classification (21) by
Cohen (2000, 2001) and the AGN type (22). A dash in column (6) indicates
that the source is not in the LW2 field. The source marked with the
label ``Var'' is a variable X-ray source detected by Brandt et
al. (2001) and below the detection threshold in more recent
observations (Brandt et al. 2001a). Sources detected in the ultra-hard
band (4-8 keV) are marked with an asterisk on the hard-band flux. For
the source \#172 we recompute I and V magnitudes on the images of
Barger et al. (1999) since the original catalogue lists one source
instead of two close sources. We classified the spectra of \#178,
\#190 and \#Var according to Cohen (2000) by inspecting the reduced
spectra of Barger et al. (1999).  Redshifts sources are: (0) Hawaii
group, (1) Cohen et al. 1996, (2) Lowental et al. 1997, (3) Hogg et
al. 1997, (4) Waddington et al. 1999, (5) Cohen et al. 2000, (6)
Dawson et al. 2001}
\label{tab:list_hdf}
\end{table*}

\begin{figure*}[!t]
\vspace*{12cm}
\caption{
Finding charts of the common X-ray and mid-IR sources in the Lockman Hole region (see Table~\ref{tab:list}). X-ray and mid-IR  isocontours are plotted on optical images (I-band) with grey and black lines, respectively.
X-ray contours come from the 0.5--7 keV image, while mid-IR contours refer to the 15$\ \mu$m image, except for sources \# 104 and \# 121 which have been detected only in the 6.75 $\ \mu$m image. The size of each image is $25''\times25''$.
}%
\label{fig:lockman_fcharts}%
\end{figure*}
\begin{figure*}[!t]
\vspace*{11.7cm}
\caption{ Finding charts of the common X-ray and mid-IR sources in the
Hubble Deep Field and Flanking Fields (see
Table~\ref{tab:list_hdf}). X-ray and mid-IR isocontours are plotted on
optical images (I-band, Barger et al. 1999) with grey and black lines,
respectively.  X-ray contours come from the 2--8 keV Chandra image,
while mid-IR contours refer to the 15$\ \mu$m image except for sources
\# 176 and \# 183 which have been detected only in the 6.75 $\mu$m
image.  The size of each image is $15''\times 15''$. Sources \# 161
and \# 175 correspond to blank fields in the I-image.  }%
\label{fig:hdf_fcharts}%
\end{figure*}

\begin{figure*}[!t]
\hbox{
\psfig{file=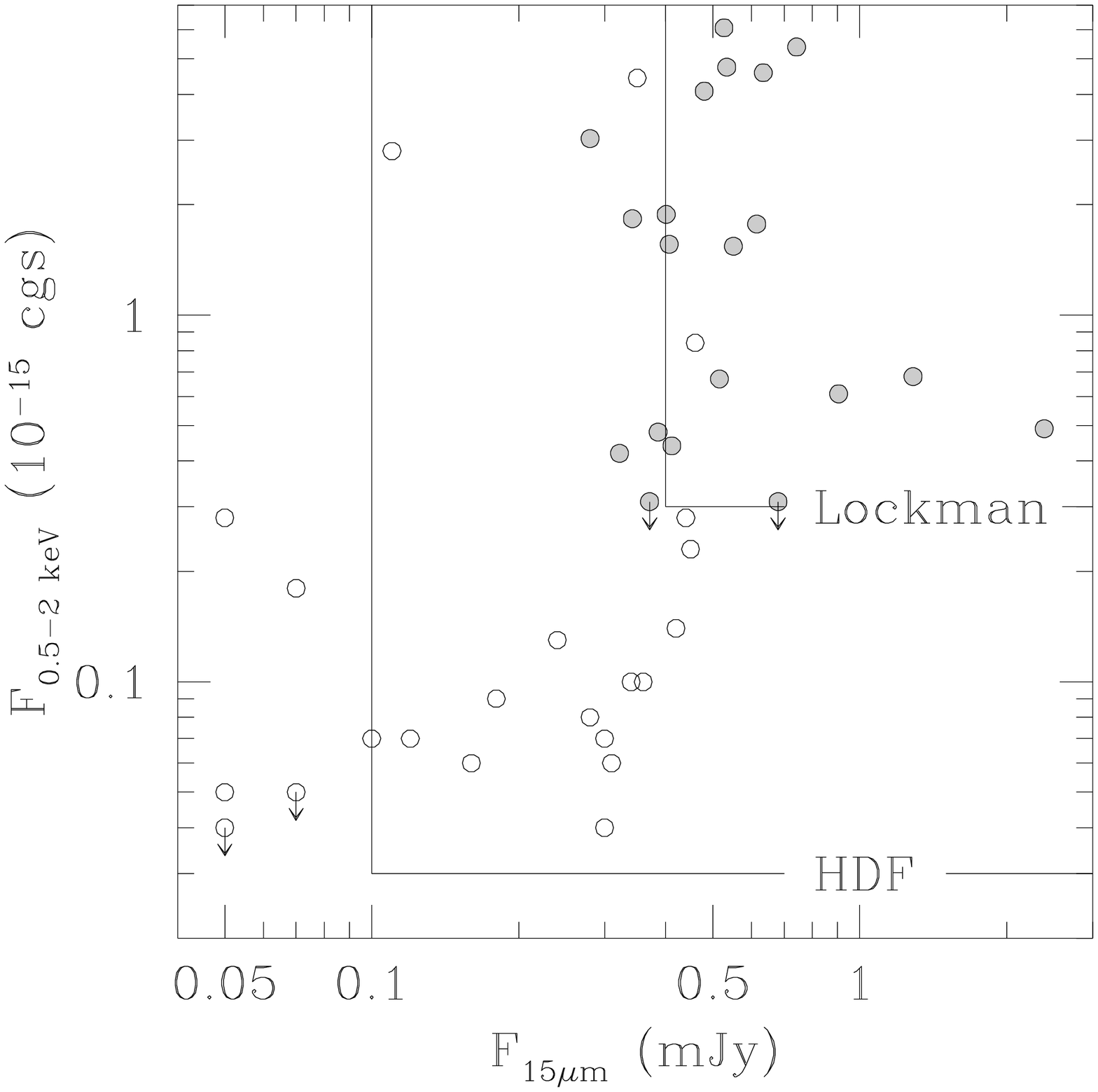,width=0.5\textwidth}
\psfig{file=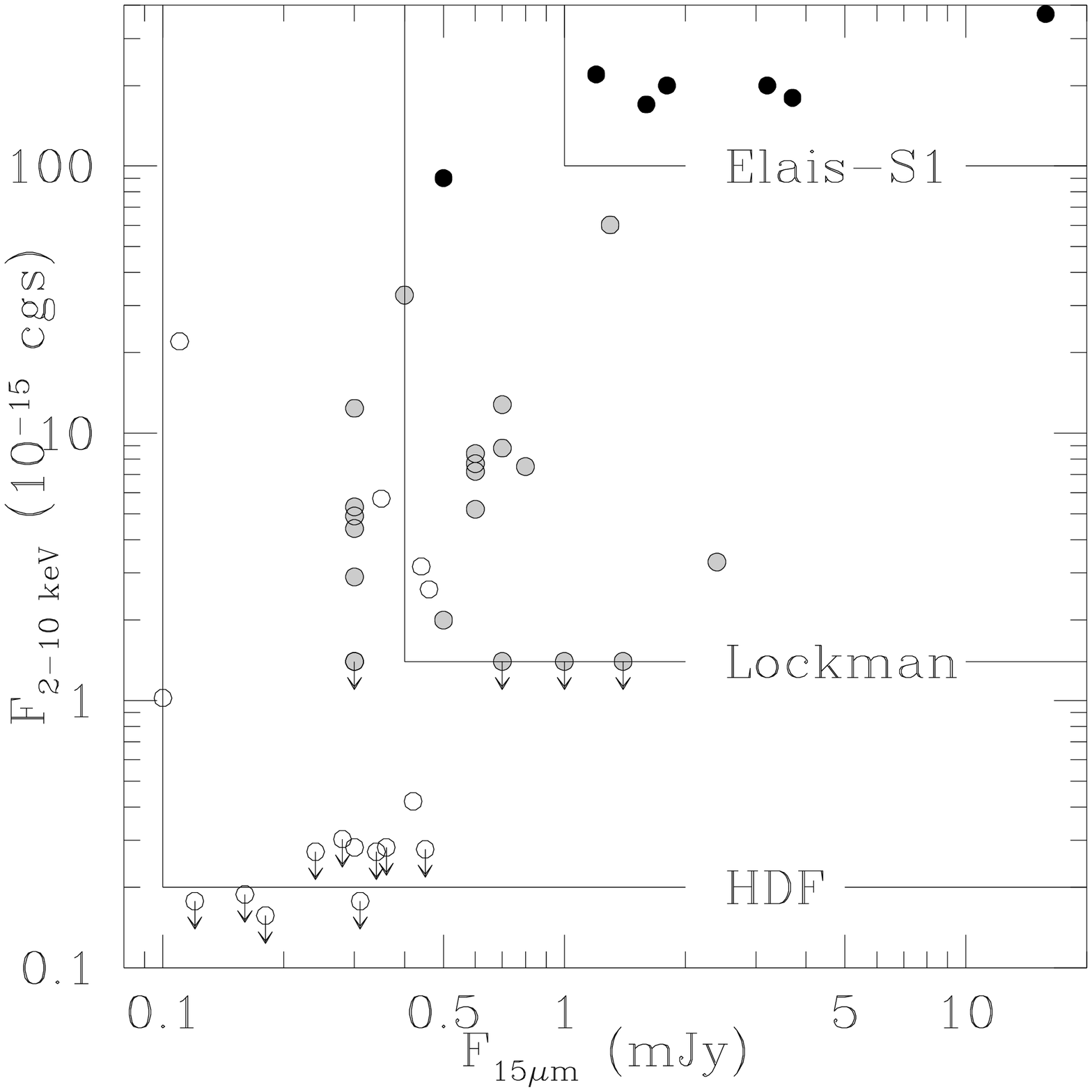,width=0.5\textwidth}
}
\caption{Mid-IR vs. X-ray fluxes for galaxies observed in the Elais-S1 (black),
Lockman (grey) and HDF (white) surveys. Diagrams refer to the soft band 
(0.5--2 keV) on the left and to the hard band (2--10 keV) on the right. The lines
delimit the X-ray sensitivity and the mid-IR (15$\ \mu$m) 80\% completeness limits.
}%
\label{fig:flux}%
\end{figure*}

\subsection{Cross-correlation of X-ray and mid-IR catalogues}

In the case of Lockman Hole observations, XMM sources have position
errors of 1--3$''$ (Hasinger et al. 2001), while 15$\ \mu$m and 6.75$\
\mu$m ISOCAM sources have position errors of 2--4$''$ (depending on the
redundancy of mid-IR observation, see Fadda et al. 2002).  In
practice, we match X-ray and mid-IR sources within a circle of 4$''$. In
total, we found 22 matches with 15$\ \mu$m sources detected at the
3$\sigma$ level and 7 with 6.75$\ \mu$m sources at the 4$\sigma$
level. For the sake of completeness, we list also upper limits of the
LW2 and LW3 fluxes when a source is detected in only one of the two
ISOCAM filters.

In the case of the Hubble Deep Field image, we have reprocessed the
ISOCAM data to recompute more accurate centres of the LW3 sources.  We
have used the I-band image of Barger et al. (1999) available on the
Web\footnote{www.ifa.hawaii.edu/~cowie/hdflank/hdflank.html} to which
we added the astrometry according to the catalogue of Hogg et
al. (2000).  Position errors of the Chandra sources are estimated less
than 1$''$, while ISOCAM sources thanks to the microscan technique of
observation have errors less than 2$''$. Therefore, we match X-ray and
mid-IR sources within a circle of 2$''$.

For each source we have computed the probability of random association
of the X-ray source with its mid-IR and optical counterparts, and
of the mid-IR source with its optical counterpart.  Assuming that the
counterpart belongs to a Poissonian distributed population of sources,
\begin{equation}
P = 1 - e^{-n(A) \pi d^2}
\end{equation}
gives the probability to have a random association
within a distance $d$ (distance between the source and the possible 
counterpart) with a source brighter than $A$ (the flux of the possible 
counterpart). $n(A)$ is the expected number of sources with
flux (magnitude) greater (lower) than that of the possible
counterpart $A$. It has been evaluated using the distribution of
mid-IR, X-ray fluxes, and the counts in selected regions of the
I-band image whitout bright stars.

Contours of the matched sources are also plotted on optical images in
Figures~\ref{fig:lockman_fcharts} and~\ref{fig:hdf_fcharts}. In most
of the cases there is a clear correspondence between X-ray and mid-IR
sources. Only in a few cases (\#149 in the Hubble Deep Field and \#41,
\#79 in the Lockman Hole) optical counterparts are uncertain and the
match relies only on the distance criterion. Three sources correspond
to blank fields, at least in these deep I-band images.

Results of these cross-correlations are reported in Tables~\ref{tab:list} 
and~\ref{tab:list_hdf} which contain positions of the X-ray sources,
distances, and probability of random associations between sources and
proposed counterparts, optical magnitudes in the V, I, R and K bands, X-ray and 
mid-IR fluxes, redshifts and AGN types when known, as well as other
quantities described in the following.

For the sources without spectroscopic redshifts, we have estimated
photometric redshifts using four optical magnitudes (V,I,R and K)
under the assumption that the optical emission is dominated by the
host galaxy. We used a library of synthetic SEDs generated with
PEGASE2.0 (Fioc \& Rocca-Volmerange 1997) to fit the distribution of
optical magnitudes. The median error on photometric redshifts, derived
from a study on the Hubble Deep Field South (Franceschini et
al., in prep.), is 0.1.

\subsection{Comparison of mid-IR and X-ray common surveys}

Up to now, only two ISOCAM surveys have been studied in the X-ray
bands: the Elais-S1 field (Alexander et al. 2001) and the HDFN
field (Hornschemeier et al. 2001).  In this paper we extend the study
of mid-IR--X-ray cross-correlation in the HDFN field
and flanking fields using the new observations of Brandt et al. (2001a)
and justifying the associations between X-ray and mid-IR sources.

Figure~\ref{fig:flux} compares the mid-IR and X-ray fluxes of the
sources detected in these surveys as well as the sensitivity limits of
the X-ray observations and the 80\% completeness limits of the 15$\
\mu$m surveys.  The survey in the Lockman Hole region is intermediate
between the Elais-S1 and HDFN surveys. It covers an area of 218
square arcminutes that is $\sim$ 30 times smaller than the Elais-S1
survey ($\sim$ 6000 square arcminutes) and $\sim$ 10 times larger than
the HDFN survey (24 square arcminutes).  In terms of sensitivity,
it is less deep than the HDFN in the soft and hard X-ray bands,
and it is definitively more sensitive than Beppo-SAX.  Moreover, the
XMM--Newton data allow us to explore with a good sensitivity the
ultra-hard energy band (4.5--10 keV), which has been pioneered by
Beppo-SAX and is not well covered by Chandra.  As we have already
seen, this band is very interesting because more than 60\% of the
ultra-hard sources in the Lockman Hole have mid-IR counterparts.

Taking into account the limits in sensitivity, the three surveys are
compatible in terms of source density.  Within the sensitivity limits
of the Lockman Hole observations ($F_{15\mu {\rm m}} \ge 0.4$mJy and
$F_{0.5-2 {\rm keV}} > 0.3\times10^{-15}$ erg cm$^{-2}$ s$^{-1}$, $F_{2-10
\rm keV} > 1.4\times10^{-15}$ erg cm$^{-2}$ s$^{-1}$) we find 13 and 11
sources in the soft and hard X-ray bands, respectively.  Therefore, we
expect to detect in the HDF within the same flux limits
1.4$^{+0.5}_{-0.4}$ and 1.2$^{+0.5}_{-0.4}$ sources in the soft and
hard band, respectively, while we detect one and two sources.
Moreover, in the hard X-ray band within the sensitivity limits of
Elais-S1, we expect to detect 0.2 $\pm$ 0.1 sources in the Lockman Hole
and 0.03 $\pm$ 0.01 sources in the HDF while no sources have been
detected in these two surveys.

In conclusion, the survey in the Lockman Hole is intermediate between
the surveys in the Elais-S1 and HDFN fields.  Due to its large
sky coverage, the Elais-S1 survey picks up very powerful and rare
hard-X ray sources. On the other hand, the deep X-ray survey in the
HDFN allows the detection of very faint X-ray sources, and therefore also normal
and starburst galaxies, in a small region of sky. So far, only the
survey in the Lockman Hole region has sufficient depth and sky
coverage to study a representative population of AGNs detected in the
ISOCAM mid-IR surveys.

\subsection{Properties of the X-ray---mid-IR matched sources}
Optical colours, redshifts and spectral classifications are available
for many of the galaxies emitting in X-ray and mid-IR bands.  In the
case of the Lockman Hole, the best known galaxies are those already
detected with ROSAT (see Lehmann et al. 2000, 2001), which constitute
approximately half of our sample.  On the contrary, redshifts are
known for all but two of the galaxies of the HDF sample due to the
great efforts made in this area (e.g.  Hogg et al. 2000; Cohen et
al. 2000), although only few galaxies are classified as AGNs or starburst 
galaxies according to their spectral features.

\subsubsection{Optical properties}
The HDFN survey, which is the deepest we consider here, allows one also to
detect  very faint sources and thus starburst and nearby galaxies
(see Hornschemeier et al. 2001; Elbaz et al. 2001). Only a small part
of the sources detected both in the mid-IR and X-rays are optically
classified as AGN (20\%).  
Also the redshift distribution of these galaxies reflects this
situation. The median redshift of 0.5 is typical of the mid-IR galaxy
population (see Fadda et al. 2002; Flores et al. 2002), while the
median redshift of the galaxies classified as AGN is 1. 
On the contrary, the sources detected in the Elais-S1 survey are almost
exclusively AGN at high redshift. 

\begin{figure*}[!t]
\hbox{
\psfig{file=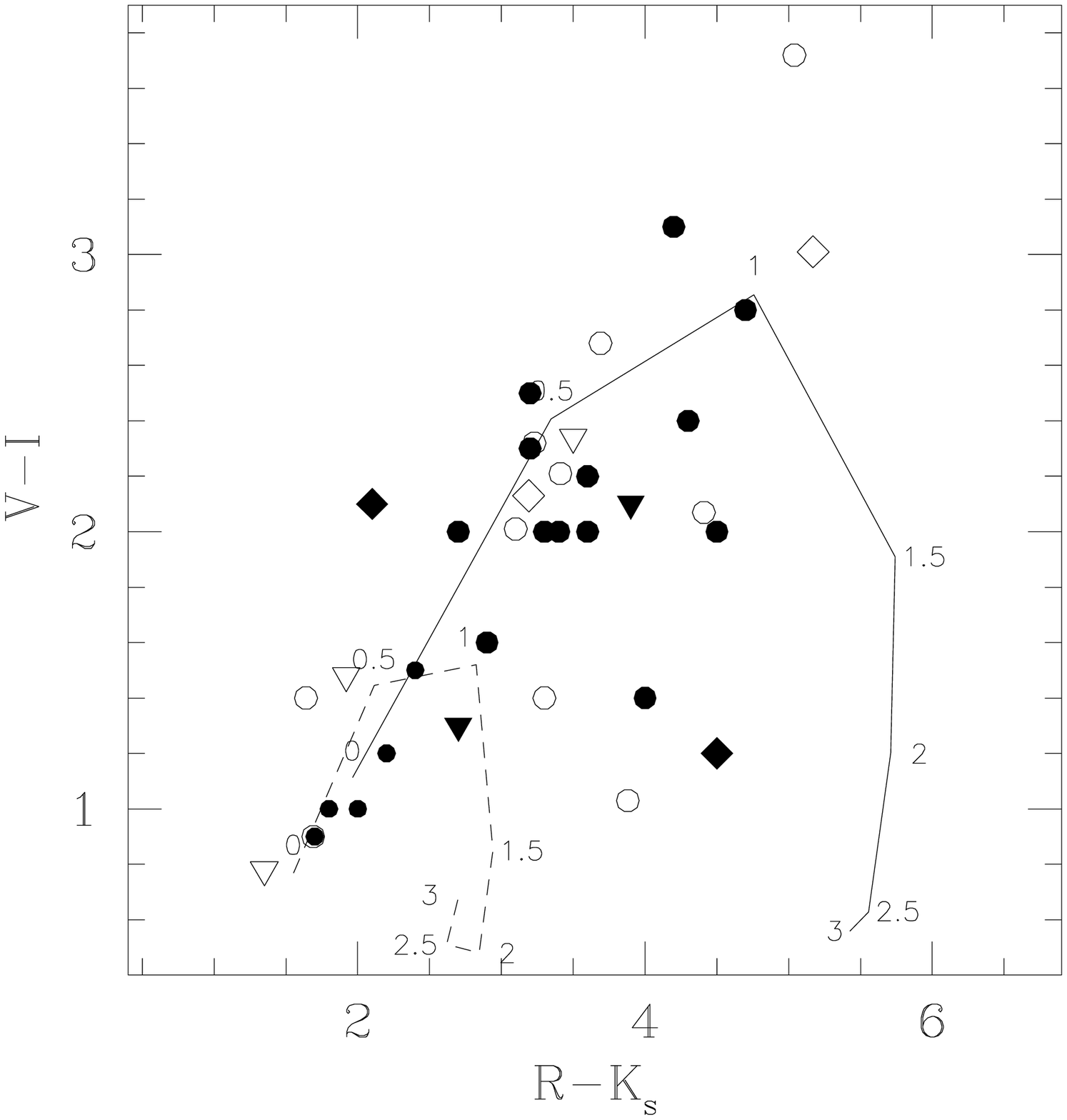,width=0.5\textwidth}
\psfig{file=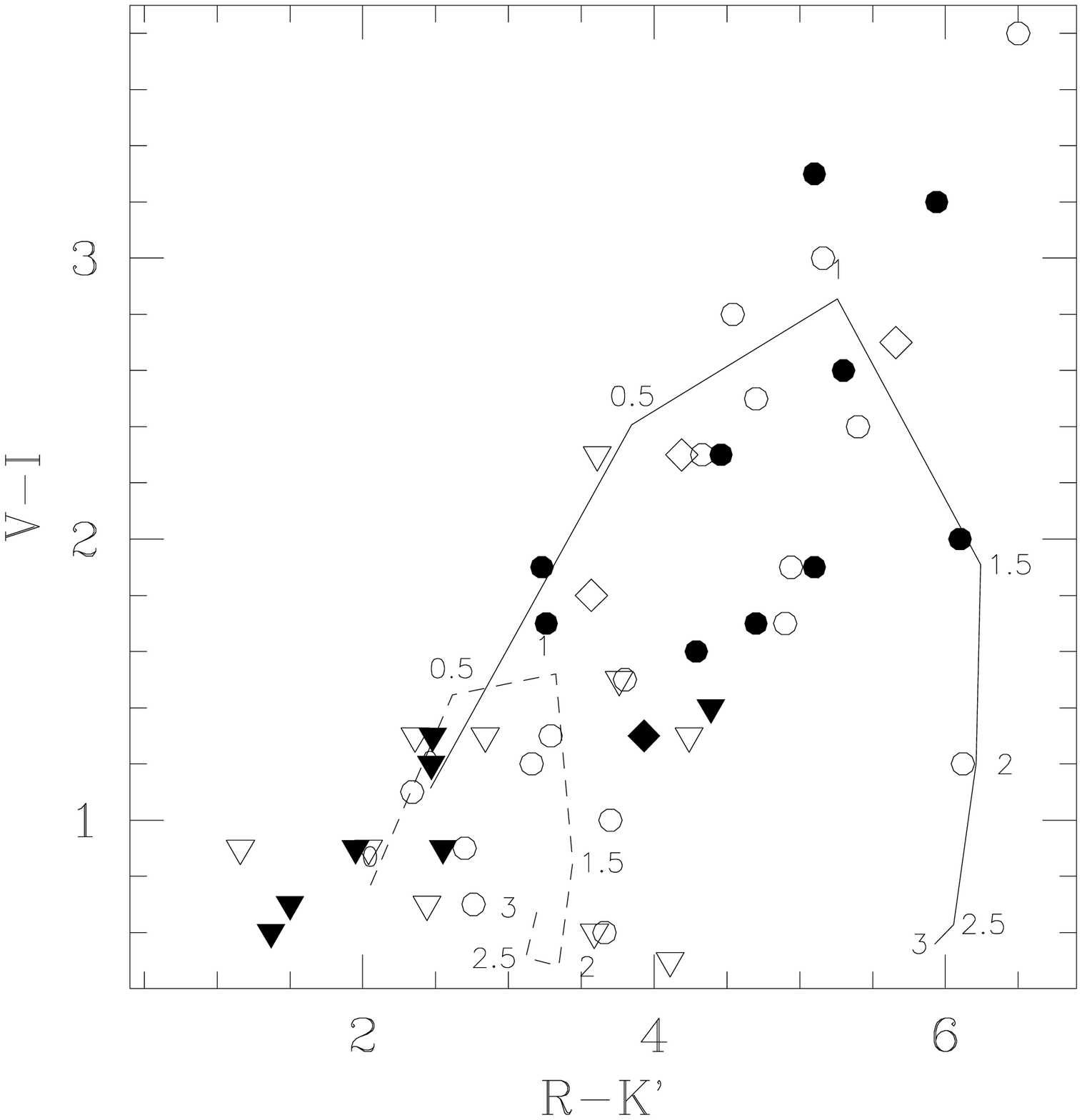,width=0.5\textwidth}
}
\caption{Optical colour diagrams for X-ray-emitting galaxies
in the part of HDFN (left) and Lockman Hole fields (right)
surveyed in X-ray and mid-IR. Triangles, diamonds and circle
refer to type-1 AGNs, type-2 AGNs and unknown type galaxies, respectively.
X-ray sources detected in the  mid-IR are marked with full symbols. 
Dashed and solid lines show colours of spiral and elliptical galaxies,
respectively, as a function of the redshift. The curves are computed
using the PEGASE2 code (Fioc \& Rocca-Volmerange 1997).
Sources with $L_{\rm X} < 10^{41}$ erg s$^-1$ in the HDF are indicated
with smaller symbols and lie around the lines of normal galaxies.
}%
\label{fig:colors}%
\end{figure*}
Excluding a normal galaxy detected
at $z=0.3$, all the other sources lie at $z>0.4$ with a median value
of $z=2$. Finally, among the sources detected in the Lockman Hole
almost half of the sample is classified as AGN while the rest is up
to now of unknown type. The median redshift of the sources is $z=1$
and all the sources lie at $z>0.4$. Therefore, this population of galaxies differs
from the bulk of the galaxies detected in the Lockman Hole, which lie
at a redshift of 0.6 (see Fadda et al. 2002).  We can learn something more
about the spectrally unclassified galaxies by looking at the optical colour
diagrams (see Figure~\ref{fig:colors}).  As expected, the type-1 AGNs
cluster in a region of blue colors while the type-2 AGNs are in
general redder and less clustered on the diagram.  Many of the
galaxies with unknown type lie in the region occupied by type-2 AGNs,
suggesting that they are highly extincted objects and probably most of
them are type-2 AGNs.

To aid in the interpretation of the diagrams, we overlay two galaxy
tracks corresponding to elliptical and spiral templates. These models
were produced with the 
PEGASE2.0 code\footnote{http://www.iap.fr/users/fioc/PEGASE.html} (Fioc \&
Rocca-Volmerange 1997) assuming a  Salpeter initial
mass function with standard cutoff (0.1--120 M$_{\sun}$). For the
elliptical track we adopt a star formation timescale of 1 Gyr,
observed at  6 Gyr, without extinction and nebular
emission. For the spiral track we consider a star formation timescale
of 5 Gyr, observed at 3 Gyr,  extinction with disk
geometry and no nebular emission.  Tracks are labelled with
representative redshifts over the range $0<z<3$, which corresponds to
the redshift range of the galaxies observed.  Few galaxies appeared
clustered around these lines. In particular, five galaxies detected in
the HDF with low X-ray luminosities have colours typical of normal
galaxies. Most of the galaxies are scattered over the diagram, but there
are almost no galaxies which follow the track of the elliptical
galaxies with $z>1$.

Finally, we note that our sample of XMM--ISO matched sources contains
five extremely obscured objects (EROs, according to the definition
$R-K \ge 5$). Objects of this type are claimed to constitute about
30\% of the optically faint X-ray sources in the deep Chandra survey
of the HDF (Alexander et al. 2001a).  On the other hand, Pierre et
al. (2001) showed that is possible to select this kind of objects
using mid-IR observations.  This sample of objects will be studied in
more detail by Franceschini et al. (2001).  In the Lockman field,
another four EROs were detected by XMM--Newton and not by ISOCAM. As
better explained in Franceschini et al. (2001), the expected 15$\
\mu$m fluxes of these objects fall below the detection limit of the
survey ($0.3$ mJy).

\subsubsection{X-ray diagnosis}

\begin{figure*}[!t]
\hbox{
\psfig{file=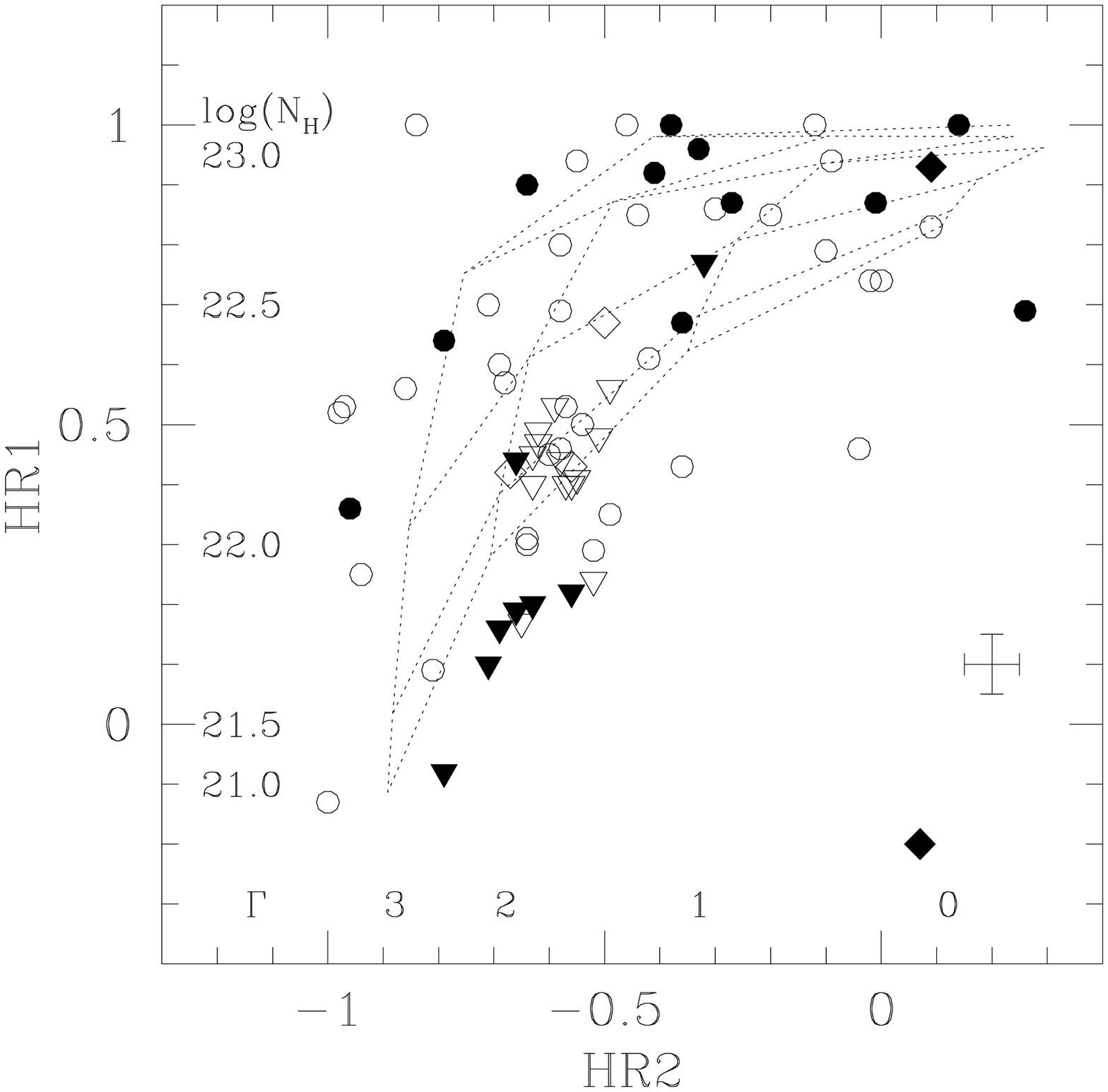,width=0.5\textwidth}
\psfig{file=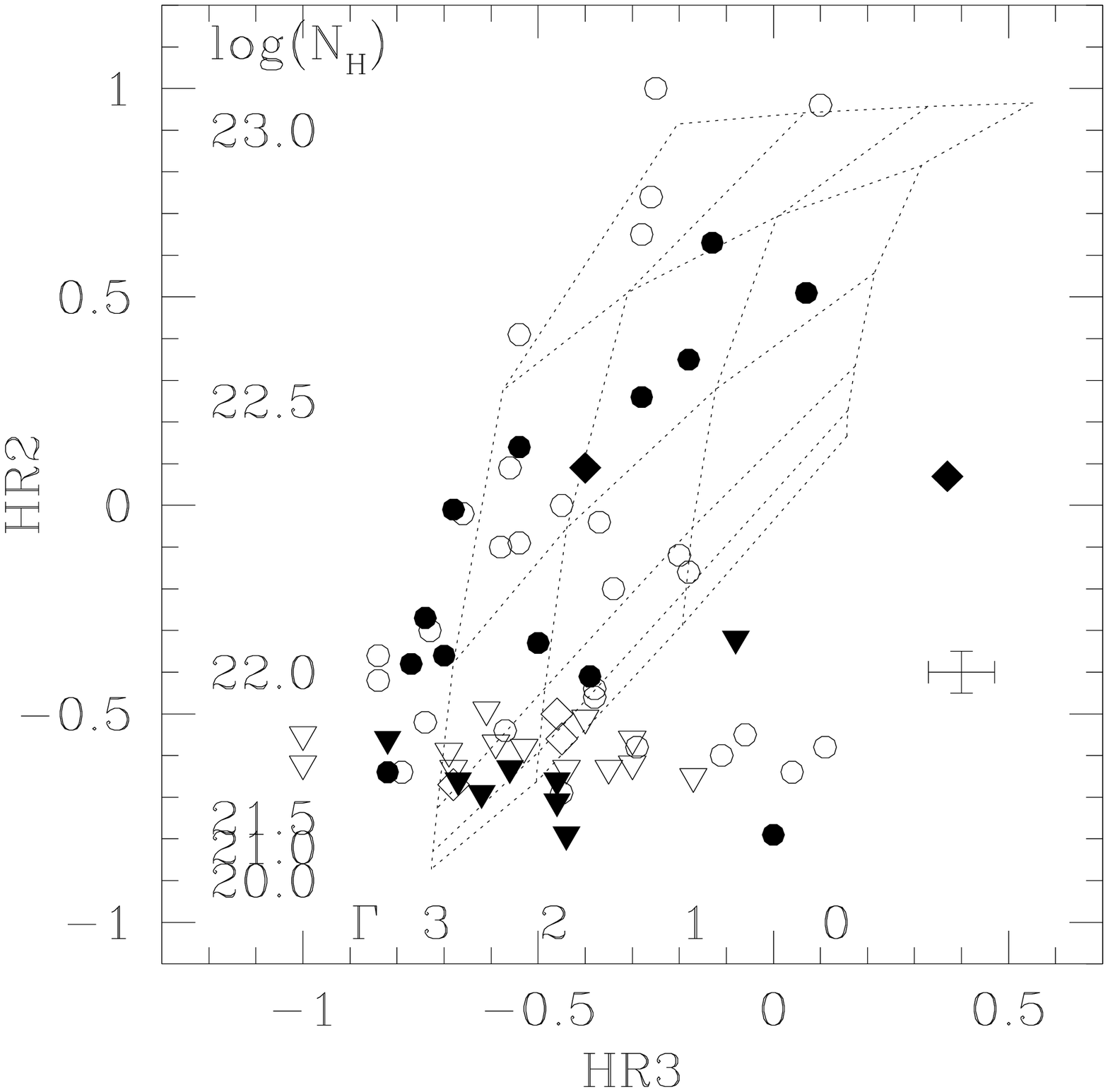,width=0.5\textwidth}
}
\caption{
X-ray diagnostic diagrams based on hardness ratios (see  Hasinger et al. 2001).
Triangles, diamonds and circle refer to type-1 AGNs, type-2 AGNs and unknown-type galaxies, respectively.
Galaxies inside the common X-ray and mid-IR area are shown. The cross indicates the
median error bar of the points. Only points with error less than 0.1 are plotted.
The galaxies with mid-IR emission are marked with full symbols. For these galaxies
HR1, HR2 and HR3 values are reported in columns 16, 17 and 18 of Table~\ref{tab:list}.
The grid gives the expected hardness ratios for power-law models with different
values of the photon index $\Gamma$ and of the neutral hydrogen absorption $\log N_{\rm H}$ (in 
the observed frame).
}%
\label{fig:hardness}%
\end{figure*}

Thanks to the large energy range which can be explored with XMM--Newton
it is possible to construct colour--colour X-ray diagrams and to classify
sources on the basis of their X-ray spectra alone (Hasinger et
al. 2001).  Figure~\ref{fig:hardness} shows X-ray spectral diagnostic
diagrams based on the hardness ratios computed using four independent
energy bands.  The hardness ratios are obtained with the formula $HR =
(H-S)/(H+S)$, where $H$ and $S$ correspond to the counts in the harder and
softer energy bands, respectively. HR1, HR2 and HR3 compare the
0.2--0.5 vs. 0.5--2 keV, 0.5--2 vs. 2-4.5 keV, and 2--4.5 vs. 4.5--10 keV
bands, respectively.  A grid representing the expected hardness ratios
for power-law models with different values of photon index ($\Gamma$)
and hydrogen absorption ($\log N_{\rm H}$) computed in the observed frame is
superimposed on the data.  The populations of type-1 and type-2 AGNs
occupy different regions in these diagrams. In particular, AGN-1
galaxies populate a limited portion of the diagrams in the soft range
(and a particularly narrow HR2 range) while the new XMM--Newton
galaxies and known AGN-2 type galaxies have harder spectra than those
of AGN-1 galaxies and occupy a larger area (see discussion in Hasinger
et al. 2001).  Also in this case, most of new XMM--Newton galaxies
detected in the mid-IR lie in a clearly separated region with respect
to the type-1 AGNs.

If we admit that unclassified galaxies are all type-2 AGNs, we detect
at 15$\ \mu$m at the 3$\sigma$ level 7 AGN-1 galaxies and 15 AGN-2
galaxies (only three of these are classified as AGN-2). This statistics
matches the percentages found in the CFRS field 1415+52 using a
multi-wavelength method to classify the galaxies.  In this case,
studying a sample of 19 ISOCAM sources, Flores et al. (1999)
classified two sources as AGN-1 and three as AGN-2. For two other
sources the classification as AGN-2 or starburst galaxies is equally
probable.

\subsection{Comparison with local templates}

Detailed spectral energy distributions (SEDs) have been obtained in
the hard X-ray band (with ASCA and Beppo-SAX) and in mid- to far-IR
(with ISO) for few local galaxies which are representative of the
classes of objects found in our samples.

Before analysing X-ray luminosities and X-ray to mid-IR spectral
indices of the galaxies of our samples, we discuss the template
galaxies which will be compared with our data.

{\em Type-1 AGNs}. Mrk509 and NGC4593 have been chosen as
typical Seyfert 1 galaxies (data from Clavel et al. 2000; Perola et
al. 2000; Guainazzi et al. 1999), while PG1613+658 has been taken as
representative of radio-quiet quasars (data from Haas et al.  2000 and
Lawson \& Turner 1997).

{\em Type-2 AGNs}. This class of object is expected to be
easily detected by combined hard X-ray and mid-IR surveys, since
almost all the UV and soft X-ray emission of the nucleus is
reprocessed into infrared light. The three templates chosen show
different behaviours.  NGC 1068, the archetypal object for the class of
Seyfert 2 galaxies (data from Sturm et al. 2000; Matt et al. 1997),
has an extremely absorbed X-ray flux. Due to this fact, it has a
mid-IR to X-ray flux spectral index which is more typical of starburst galaxies
than type-2 AGNs (see Figure~\ref{fig:alpha}).  We also chose the
Circinus galaxy, a Seyfert 2 object with a reflection-dominated
spectrum in the 2--10 keV range and a transmitted component above 10
keV (data from Sturm et al. 2000; Matt et al. 1999; Sambruna et al.
2001). Finally, we
report the Sy2 NGC6240 (data from Charmandaris et al. 1999; Vignati et
al. 1999) whose energy output, according to Vignati et al. (1999) is
dominated by the AGN and not from star formation, as deduced by Genzel
et al. (1998) on the basis of the ISO spectrum.

{\em Starbursts}. M82 and NGC253, two of the nearest starburst galaxies,
are assumed as typical templates for galaxies with active star formation
(data from Sturm et al. 2000; Cappi et al. 1999).

{\em Ultraluminous galaxies}. Galaxies of this class, which emit large
parts of their bolometric luminosity in the infrared, are known to be
 powered mainly by star formation, although a small fraction of
the emission is probably due to AGN activity (e.g. Lutz et al. 1998;
Tran et al. 2001).  We chose Arp 220 as an example of an ultraluminous
starburst galaxy (data from Sturm et al. 1996; Charmandaris et
al. 1999; Iwasawa et al. 2001).

\begin{figure*}[!t]
\hbox{
\psfig{file=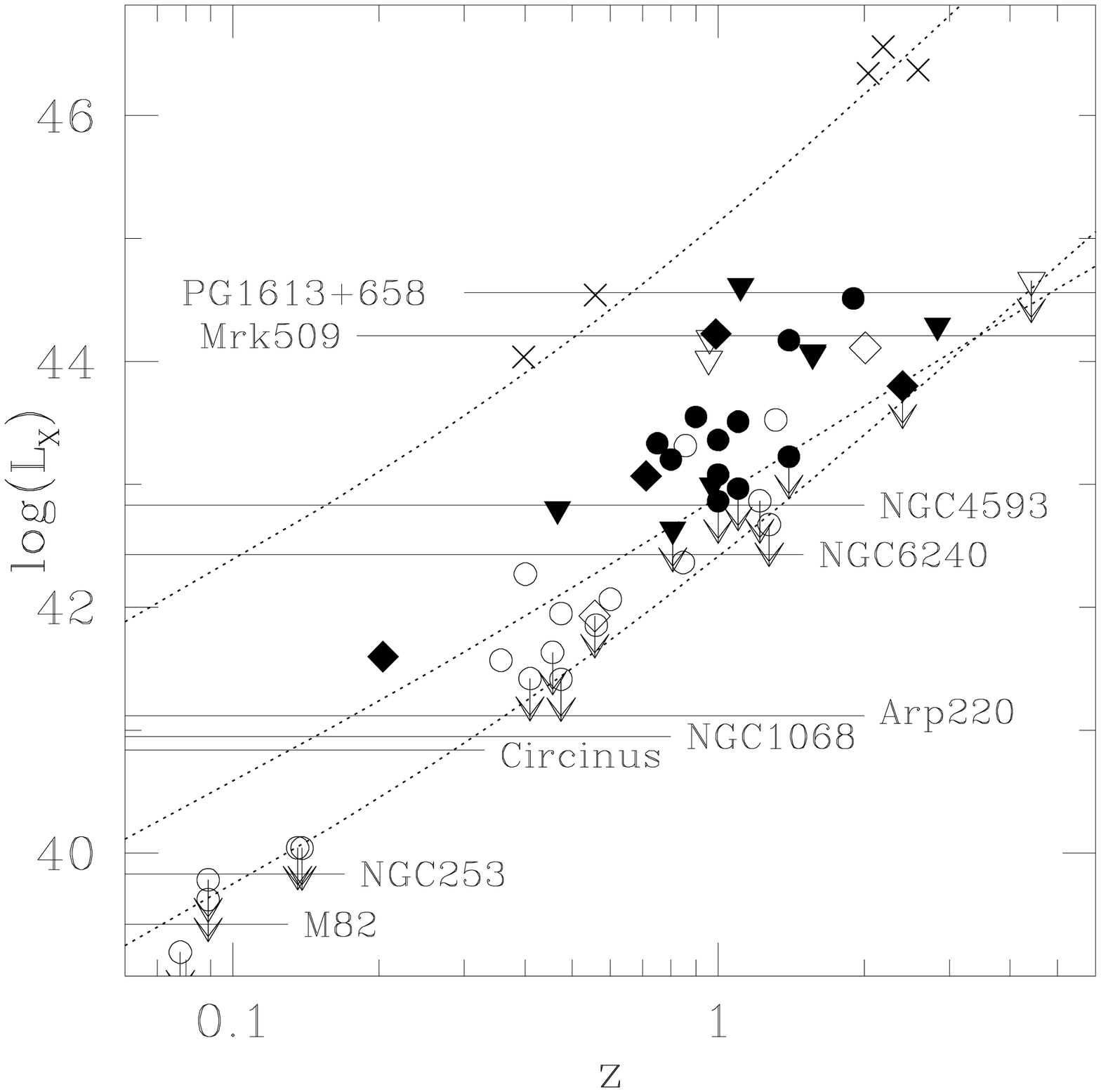,width=0.5\textwidth,angle=0}
\psfig{file=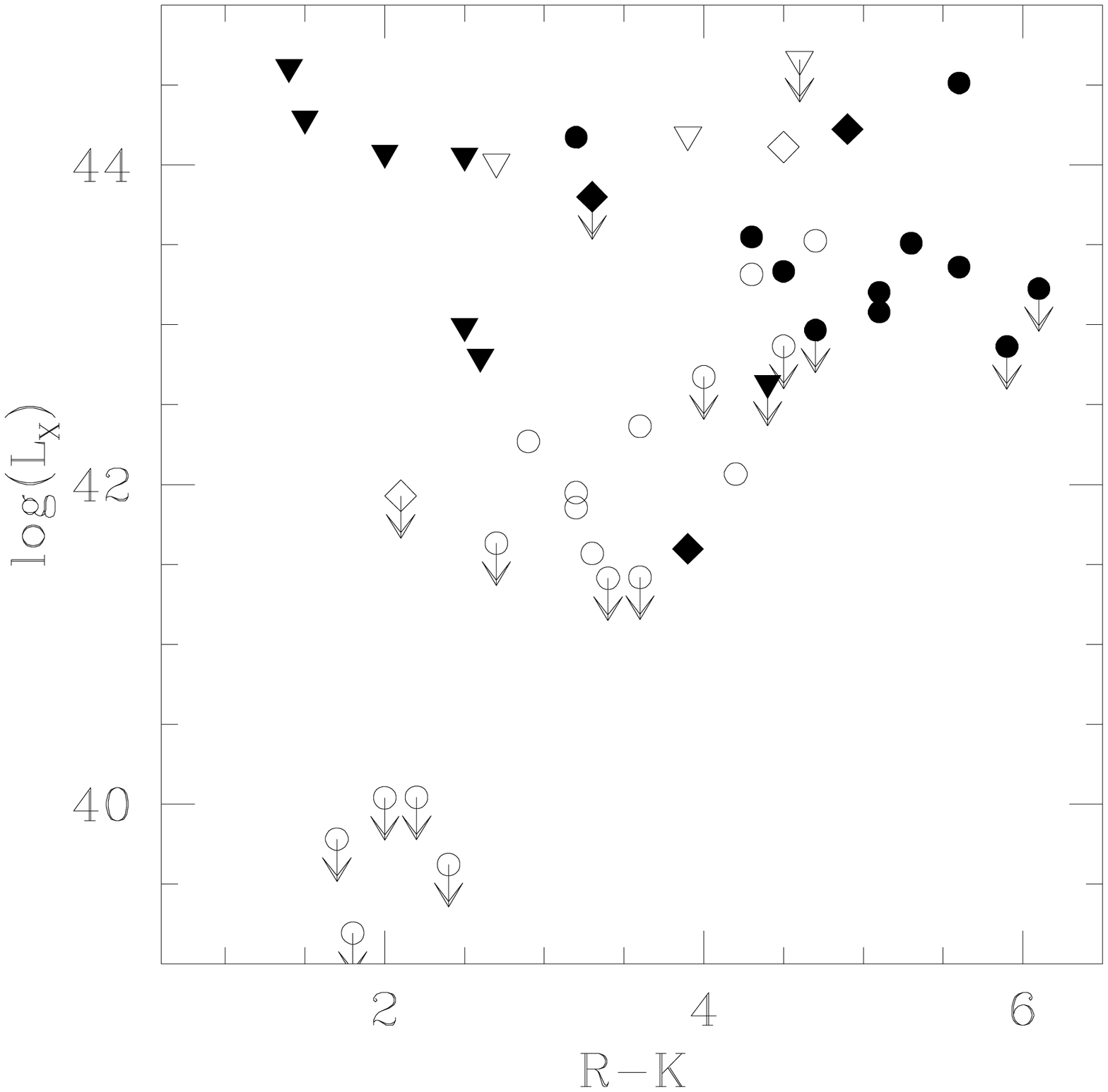,width=0.5\textwidth,angle=0}
}
\caption{2--10 keV X-ray luminosity versus redshift (left) and R $-$ K colour (right) for the
X-ray mid-IR matched sources. Open symbols, grey symbols and crosses refer
to the HDF, Lockman and Elais surveys, respectively.
Type-1  and type-2 AGNs are marked with triangles and squares, respectively,
while circles identify unclassified sources.
In the left figure, the three dashed lines show the sensitivity limits of the
three X-ray surveys. The horizontal lines trace the X-ray luminosity of the template
galaxies discussed in the text. The luminosity distance is computed according
to Carroll, Press \& Turner (1992) assuming a Universe with $H_0$ = 70 km s$^{-1}$ Mpc$^{-1}$,
$\Omega_\lambda$=0.7 and $\Omega_M$=0.3.
}%
\label{fig:lum}%
\end{figure*}

\subsubsection{X-ray luminosities}

Since we have spectroscopic and photometric redshifts for almost all the galaxies
of our samples, it is possible to compute X-ray luminosities of these galaxies
and compare them with those of local templates.

To compute the luminosity distance we assume an Universe with $H_0$ = 70
km s$^{-1}$ Mpc$^{-1}$, $\Omega_\lambda$=0.7, and $\Omega_M$ = 0.3 using
the formula in Carroll, Press \& Turner (1992).
As we can note in Figure~\ref{fig:lum}, the sources in the Lockman Hole
sample have 2--10 keV luminosities between 10$^{42.5}$ erg s$^{-1}$ and
10$^{45}$ erg s$^{-1}$, which are typical of luminous type-2 AGNs and
normal type-1 AGNs. Sources detected in the Elais-S1 have luminosities
typical of type-1 AGNs. Finally, among the sources detected by Chandra
in  HDFN, we find low luminosities sources at low redshift
which are probably starburst galaxies, few galaxies in the luminosity
range populated by Lockman sources, and a population of galaxies with
intermediate luminosities which could be ultraluminous infrared galaxies
or low-luminosity type-2 AGNs.

In the same  Figure~\ref{fig:lum} we plot also the hard X-ray luminosity 
versus the R $-$ K colour. This allows one clearly to  segregate normal galaxies
which are faint X-ray sources and $R-K \sim 2$ and type-1 AGNs, which populate
the left upper corner of the diagram. It is still difficult to distinguish
type-2 AGNs from ultraluminous galaxies.

\subsubsection{Mid-IR to X-ray spectral index}

A way to combine the information coming from mid-IR and X-ray fluxes is
to compute the mid-IR to X-ray spectral index $\alpha_{\rm IX}$ assuming a power law
spectral energy distribution: $F_{\nu}\propto \nu^{-\alpha_{\rm IX}}$.
Values reported in Tables~\ref{tab:list} and~\ref{tab:list_hdf}
are computed  using the observer frame flux densities at 15$\
\mu$m and 5~keV. The flux densities at 5 keV have been derived from
the observed 2--10 keV fluxes (2--8 keV in the case of HDFN Chandra
data) and spectral indices.

\begin{figure*}[!t]
\hbox{
\psfig{file=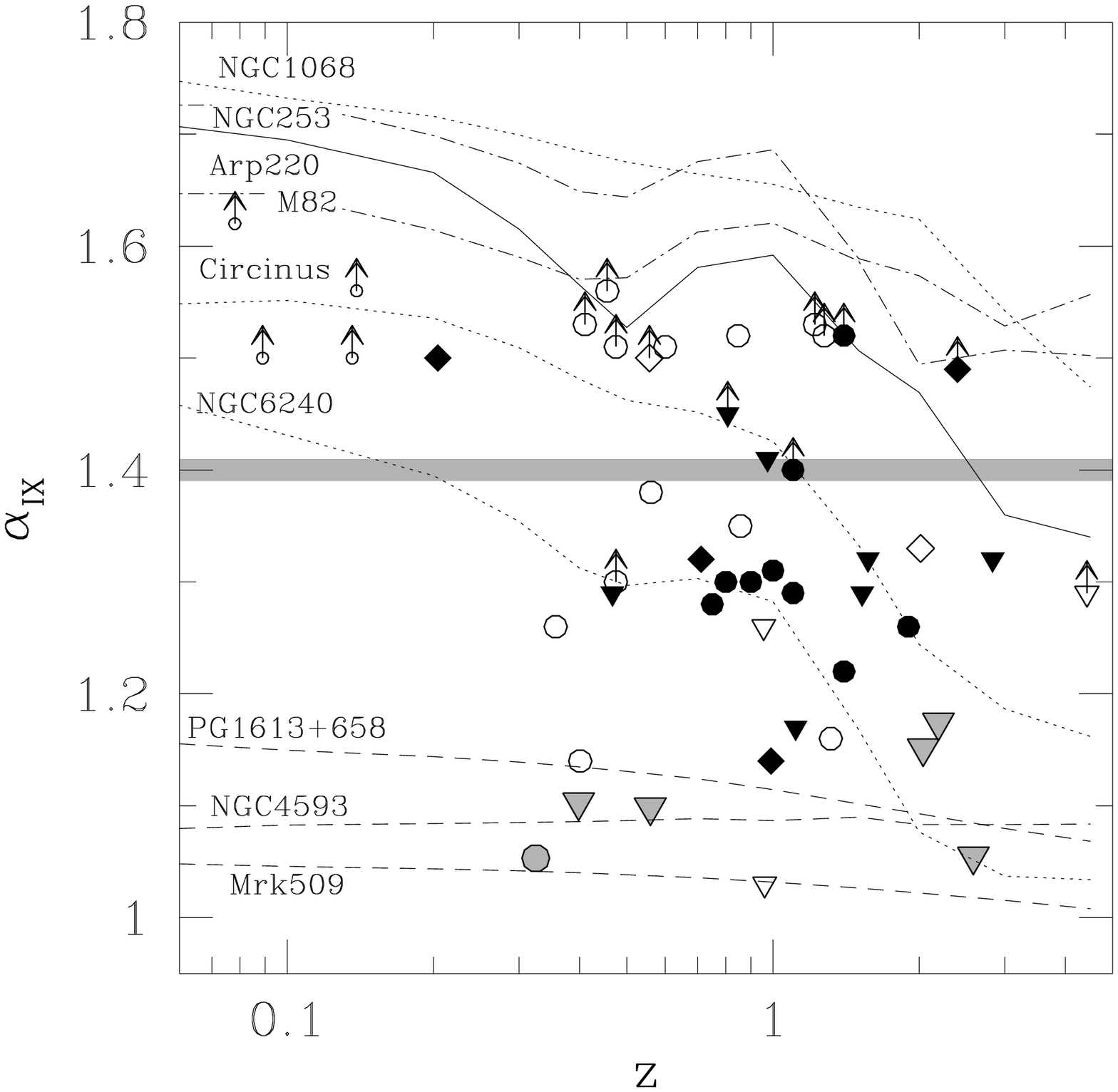,width=0.5\textwidth}
\psfig{file=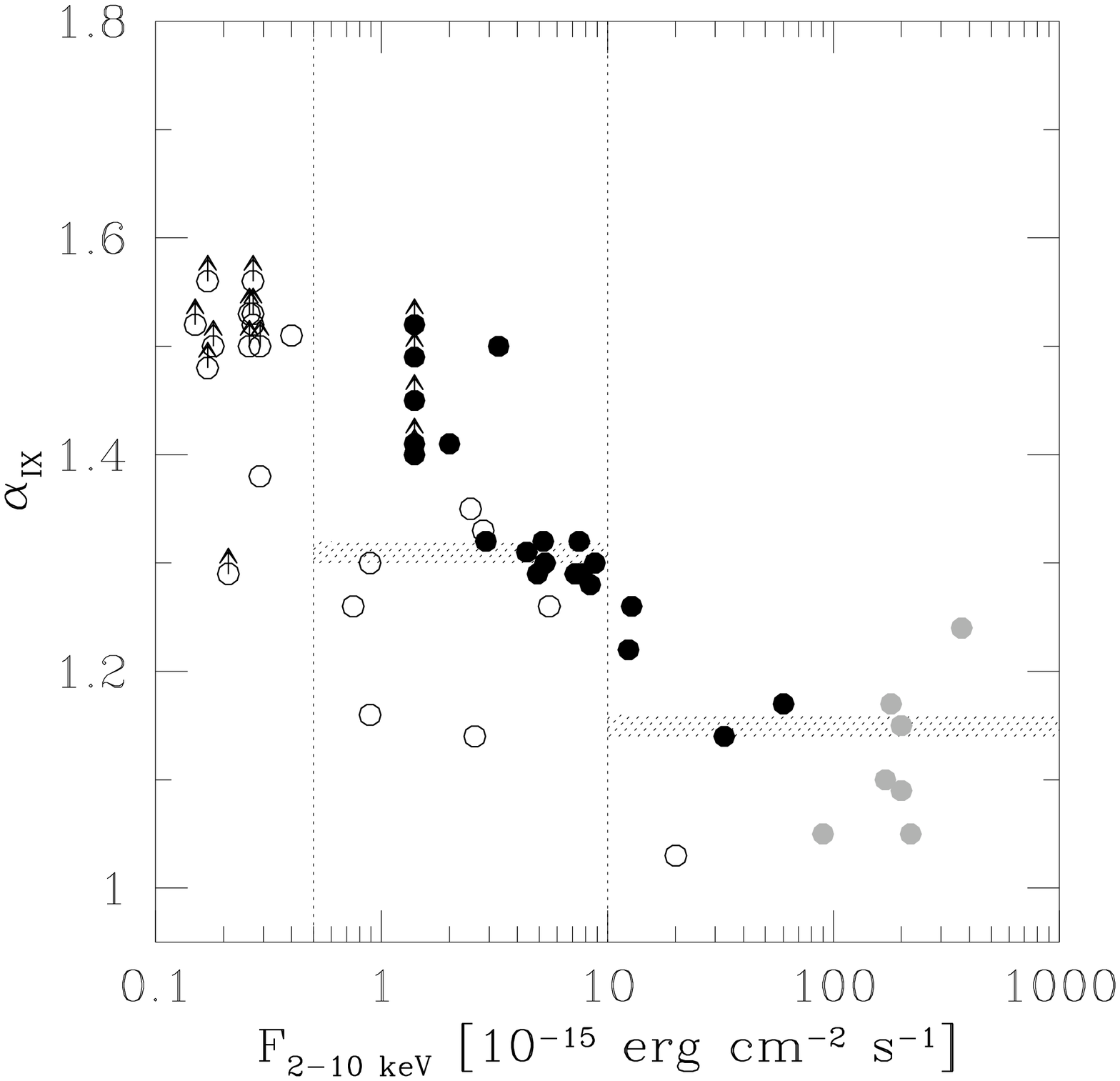,width=0.5\textwidth}
}
\caption{ On the left: distribution of the observed mid-IR to hard
X-ray spectral indices $\alpha_{\rm IX}$ as a function of redshift. Open,
grey and black symbols refer to objects detected in the HDFN,
Elais-S1 and Lockman Hole surveys, respectively. Triangles, diamonds
and circles represent type-1 AGNs, type-2 AGNs and unknown types,
respectively. HDF galaxies with $L_{\rm X} \le 10^{40}$ erg s$^{-1}$ are drawn with small
symbols.  Templates derived from various types of known active and
starburst galaxies are shown (see the text). The horizontal band shows
the $\alpha_{\rm IX}$ of the cosmic background.
On the right: $\alpha_{\rm IX}$ versus the 2--10 keV flux. Median 
values in the flux ranges assumed to compute the AGN contribution
to the mid-IR background are drawn.
}%
\label{fig:alpha}%
\end{figure*}

In Figure~\ref{fig:alpha}, which gives $\alpha_{\rm IX}$ as a function of
redshift, we report all the sources detected in the Lockman, HDFN
and Elais-S1 surveys.  We also show the values of $\alpha_{\rm IX}$ as a
function of redshift for the aforementioned local templates.

Galaxies dominated by star formation (starburst and ultraluminous galaxies)
have high values of $\alpha_{\rm IX}$ at any redshift. On the contrary, type-1
AGNs have quasi-constant values between 1 and 1.2. Between these two
envelopes of curves we find the templates of type-2 AGN Circinus and 
NGC 6240. Only NGC 1068, which is known to have an atypical SED having,
for example, a flat mid- to far-IR spectra (see Elbaz et al. 2001), lies in
a region of the diagram occupied by starburst-dominated galaxies.

Most of the galaxies detected in the Lockman Hole survey populate the
region of the diagram delimited by type-2 AGNs (Circinus and NGC 6240).
The galaxies detected in the Elais-S1 surveys lie around the type-1
AGN curves.

The HDFN survey, due to its high sensitivity, is able to detect
also non-active galaxies with high $\alpha_{\rm IX}$ index. In fact, half
of the HDF sources lie just below the curves of starburst and
ultraluminous galaxies, while the other half have $\alpha<1.4$.
Combined with the information on the X-ray luminosity, we will use
this diagram to discriminate between HDF sources whose emission is
dominated by AGN or star formation activity.

 It is interesting to remark that a large part of the type-1 AGNs
detected in the Lockman area have an $\alpha_{\rm IX}$ index greater than
those of the local templates. Except for one case which is an absorbed
type-1 AGN, as revealed by the X-ray hardness ratio diagrams (\# 79
in Table~\ref{tab:list}), the most probable explanation is that star
formation of the host galaxies contribute a large fraction of the
mid-IR flux. Hence, their $\alpha_{\rm IX}$ values should differ
significantly from those of local templates, for which we can easily
discriminate between the host galaxy and AGN.

\subsection{Contribution to the extragalactic background light}
 
The samples of sources discussed allow us to estimate how much of the
mid-IR extragalactic light detected in the mid-IR surveys is due to
AGNs. We can derive this quantity in a direct way by simply computing
the total of the mid-IR fluxes of the sources whose emission is
dominated by AGNs and dividing this by the total of the mid-IR fluxes of
the sources in the area. In this case, we can  estimate only the AGN
contribution within the sensitivity limits of the surveys and not to
the total extragalactic background mid-IR light.

Alternatively, we can use the median $\alpha_{\rm IX}$ for different classes
of contributors to the X-ray background to estimate the total
contribution to the mid-IR extragalactic background (following 
Severgnini et al. 2000).

\subsubsection{Estimating the AGN contribution to mid-IR surveys}

We can derive the AGN contribution in the case of HDF and Lockman 
surveys for which we have the complete information on X-ray and mid-IR
sources. 
To do this, we have to select on the basis of the optical, X-ray
and mid-IR properties, the subsamples of sources whose mid-IR emission
is dominated by AGNs.

In the case of the Lockman Hole, we have seen that almost all the galaxies
in the sample have high X-ray luminosity and low $\alpha_{\rm IX}$ values.
Therefore, we conservatively assume that the mid-IR emission
of all the galaxies in the sample  is due to AGNs.

\begin{figure}[!bh]
\vspace*{-1cm}
\psfig{file=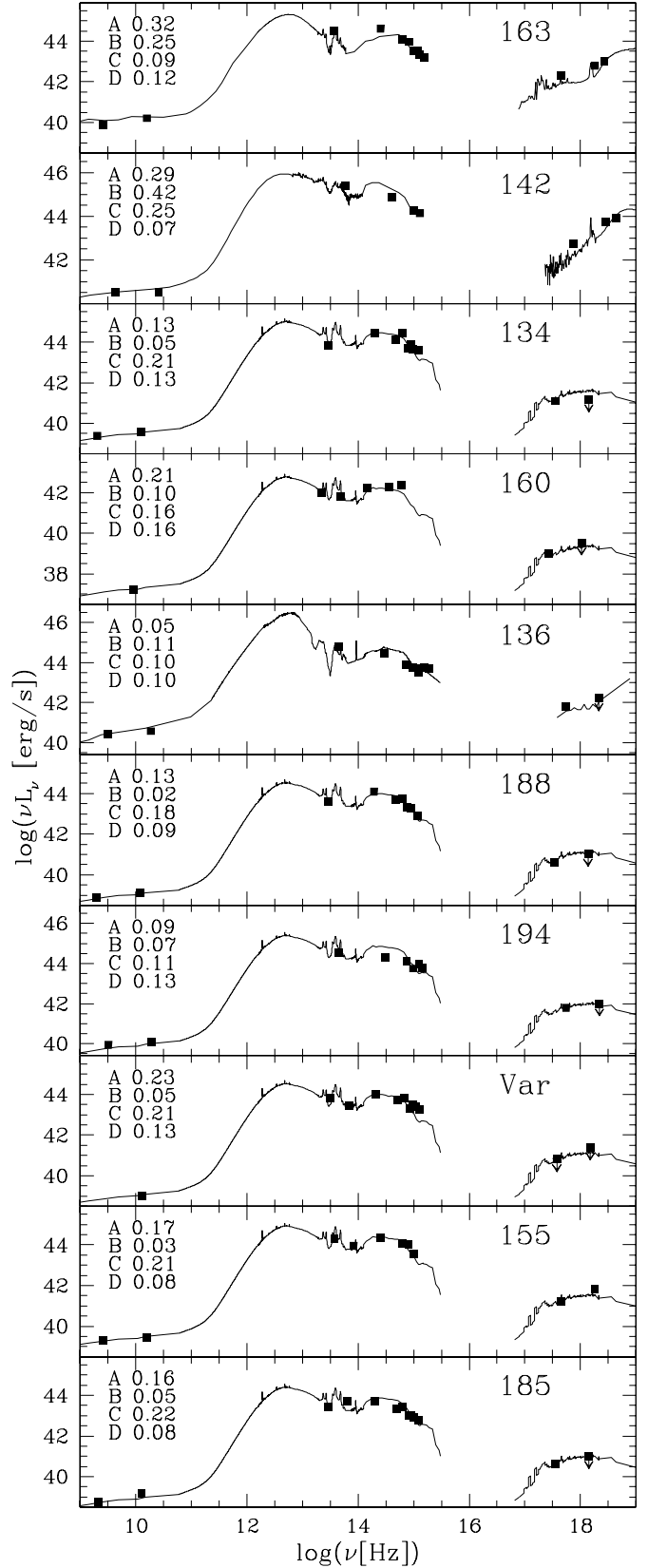,width=0.5\textwidth}
\caption{Radio, mid-IR, near-IR, optical and X-ray data for sources
in the HDFN (numbered as in Table~\ref{tab:list_hdf}) superimposed to scaled SEDs of template galaxies (A: Arp220, B: M82, C: NGC6240, D: Circinus).
The fit with the lowest $\chi^2$ (see values in the upper left corner) is shown.}%
\label{fig:seds}%
\end{figure}

In the case of the HDF, we have seen that Chandra observations are so
deep that also the X-ray emission coming from starburst galaxies is
detected. Thus, in order to estimate the AGN contribution to the mid-IR
total emission we have to select the galaxies whose mid-IR emission
is dominated by the AGN. We base our selection on the X-ray luminosity (see
Figure~\ref{fig:lum}) and on the shape of the SED from radio to X-ray
wavelengths (see Figure~\ref{fig:seds}). 

Out of 16 sources with flux greater than 0.1 mJy, four have a high
X-ray luminosity ($L_{\rm X} > 10^{43}$ erg s$^{-1}$) and another four are
faint in the X-ray ($L_{\rm X} \le 10^{40}$ erg s$^{-1}$). In these cases,
we assume that the mid-IR emission is dominated by AGN and
star-formation activity, respectively. Moreover, since the source \#172
is detected also in the ultra-hard band, we consider that it is
dominated by AGN activity. We classify all the other sources, which
have an intermediate X-ray luminosity, by comparing their radio,
mid-IR, near-IR, optical and X-ray data with the SEDs of two
star formation-dominated and two type-2 AGNs (Arp 220, M82, Circinus
and NGC 6240) for which we have the SED from radio to X-ray frequences.
Radio data at 8.5 GHz come from Richards et al. (1998) and at 1.4 GHz
from Richards (2000). In the 1.4 Ghz case, we also retrieved  the
image\footnote{http://www.cv.nrao.edu/~jkempner/vla-hdf/} to estimate
the 1.4 GHz flux of the source \#155. 
For each galaxy we fitted the data to the template SEDs scaled in luminosity
choosing the fit with the lowest $\chi^2$ value. Figure~\ref{fig:seds} shows
the $\chi^2$ values for each SED and the the best fit superimposed to  the
data. In the fit we considered also  5 keV  upper limits. This means that
the $\chi^2$ of the NGC6240 and Circinus SEDs are typically underestimated.
For comparison, in the same figure,  the SED of two bright
X-ray sources (\#163 and \#142) and one faint X-ray source (\# 160)
are also shown (the best fit is obtained with the NGC 6240, Circinus and M82
SEDs, respectively). The sources with intermediate X-ray luminosity
are all well fitted with the M82 SED, except for \# 136 which is fitted
by the Arp 220 SED. We note that also the variable X-ray source (``Var'')
follows the M82 SED very well.
Therefore, in these cases we assume that the mid-IR emission of
these sources is not dominated by the presence of an AGN.

\begin{figure*}[!t]
\hbox{
\psfig{file=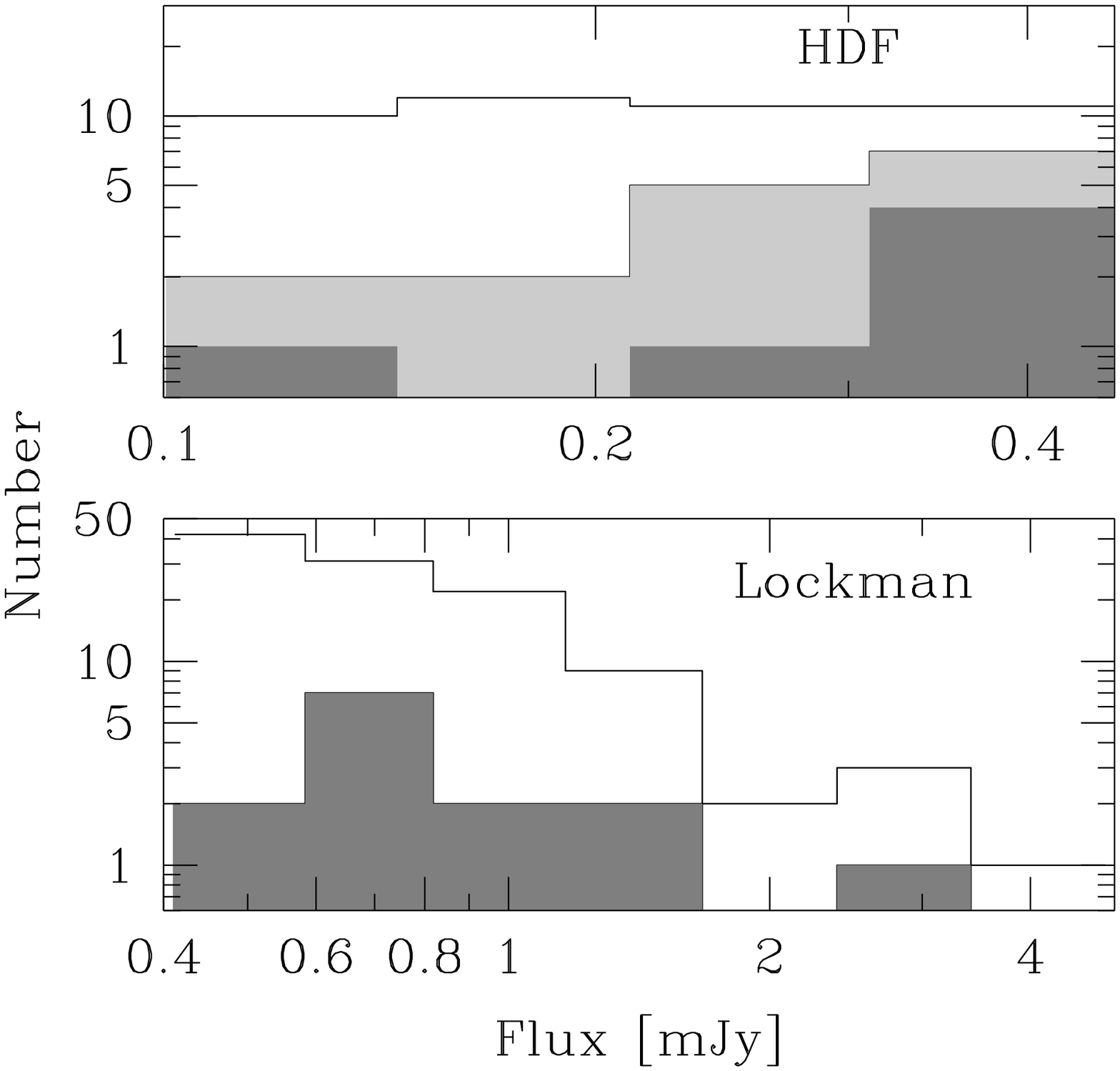,width=0.5\textwidth}
\psfig{file=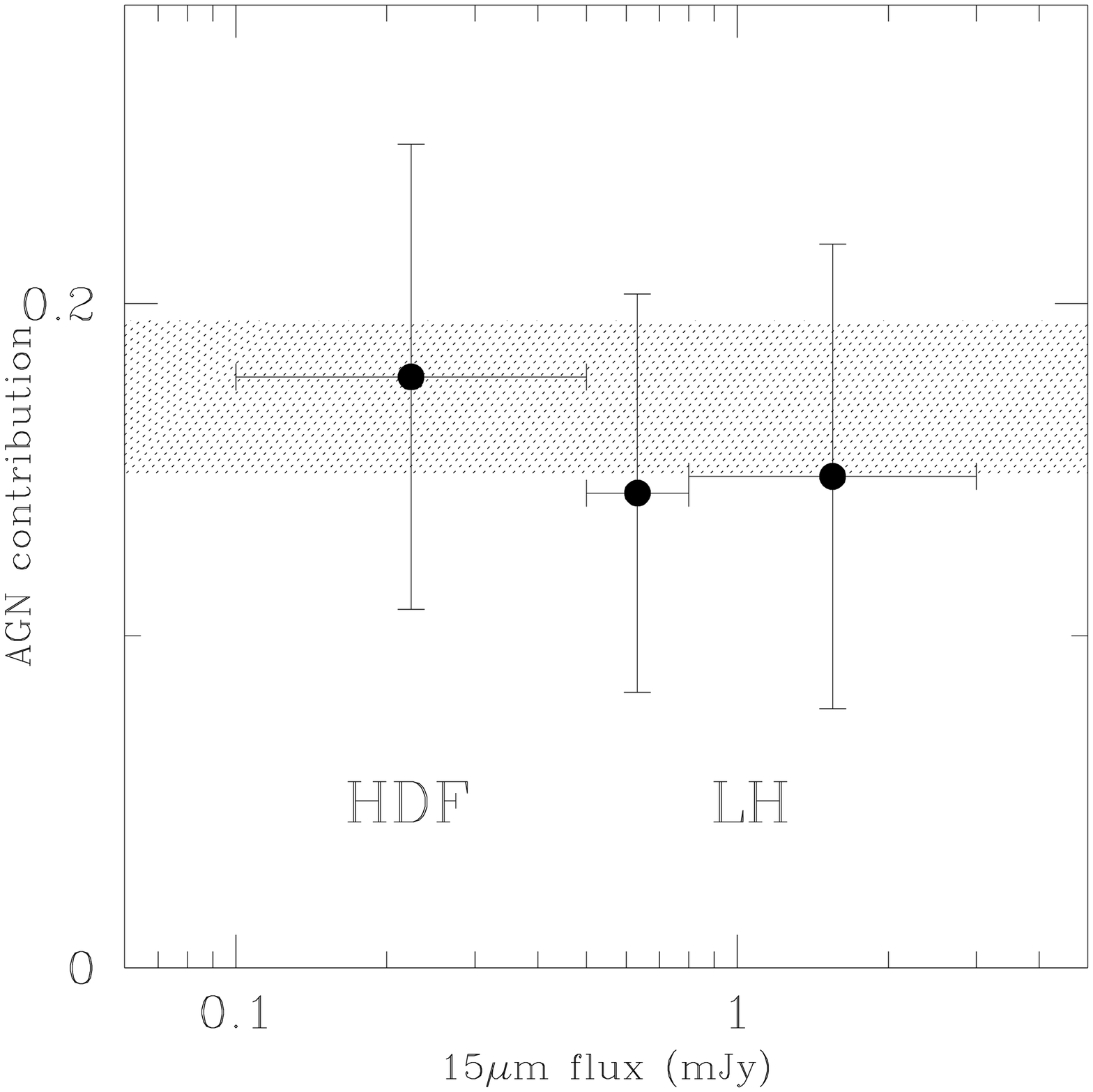,width=0.5\textwidth}
}
\caption{ On the left: histogram of LW3 fluxes in the Lockman and HDF
surveys. Sources detected in the X-ray are shaded, while the black histogram
shows the sources dominated by AGN emission. On the right:
ratio of integrated 15$\ \mu$m flux of AGN-dominated sources
to that of all the mid-IR extragalactic sources as a function of
limiting flux. The 0.1 mJy points comes from the HDFN survey, while
the other points are computed on the basis of the Lockman survey.
The horizontal band shows the percentage
of background light due to AGN emission according to the analysis
based on the  $\alpha_{\rm IX}$ index (see the text).}%
\label{fig:agnratio}%
\end{figure*}

In Fig.~\ref{fig:agnratio} we summarise the contribution of AGNs to the
mid-IR extragalactic background as a function of the flux. The bin
between 0.1 mJy and 0.5 mJy has been defined using the HDF data, since
the sources detected in this field cover  this range of fluxes well and
the HDF is complete for fluxes greater than 0.1 mJy (Aussel et
al. 1999). Due to its small size, there are no sources in the HDF with
fluxes greater than 0.5 mJy. Hence, the contribution in the  other two
bins is based on the Lockman Hole data which are more than 80\%
complete at the flux of 0.5 mJy (Fadda et al. 2002).

In the HDF there are 42 sources for a total of 9.9 mJy in the 0.1--0.5 mJy
bin, five of which are classified by us as AGN-dominated. This implies
an AGN contribution in this flux bin of $(17.8 \pm 7)\%$.

The Lockman Hole survey covers well the 0.5--3 mJy flux interval where
we find 103 sources for a total of 81.8 mJy with 13 sources which are
AGN-dominated, leading to a total contribution of $(14.6 \pm 4.7)\%$.
In Fig.~\ref{fig:agnratio}, we report the contribution in two bins:
0.5--0.8 mJy ($(14.3 \pm 6)\%$) and 0.8--3 mJy ($(14.8 \pm 7)\%$).
The contribution in this interval is probably slightly underestimated
because, as  is clear from Fig.~\ref{fig:lum}, XMM--Newton
observations miss a population of fainter X-ray sources which probably
contain highly obscured AGNs. The effect should not be dramatic because,
as we have seen in our analysis of the HDF sources, the mid-IR emission
of most of these intermediate X-ray luminous sources is not dominated by
the AGN activity.

From these estimates, we can derive the AGN contribution to the
fraction of the mid-IR extragalactic background due to the emission
of 0.1--3 mJy sources, which constitute $\sim$ 70\% of the measured
background.  The 0.1--0.5 mJy and 0.5--3 mJy sources contribute 48\%
and 23\% of the observed mid-IR background, respectively (Elbaz et
al. 2001).  Therefore, AGNs contribute $(16.8\pm6.2)\%$ of the
fraction of the mid-IR extragalactic background for which are responsible
the sources detected in the 0.1--3 mJy flux interval.

\subsubsection{Estimating AGN contribution with median mid-IR to X-ray spectral indices}

Following Severgnini et al. (2000), we can estimate the mid-IR
contribution of the X-ray sources using the mean $\alpha_{\rm IX}$ indices
of bright and faint X-ray contributors to the X-ray background with
mid-IR emission.

To apply this method we have to know the values of the X-ray and
mid-IR backgrounds. We compute the 5 keV X-ray background using the
estimation of the 1--7 keV background by Chen et al. (1997), which is in
good agreement with recent Chandra and XMM counts. In particular, the
counts by Brandt et al. (2001a) clearly flatten at low fluxes,
indicating that almost all the background is resolved in this
survey. Assuming the background of Chen et al. (1997), Alexander et
al. (2001a) evaluate that $\sim 86\%$ of the 2--8 keV background is
resolved by Chandra observations in the HDF region. The recent
estimation by Vecchi et al. (1999) with Beppo-SAX observations seems
to be too high to agree with recent deep observations of XMM and
Chandra satellites. In the case of 15$\ \mu$m, the total background
has not yet been measured.  Observational values are the upper limit
of 5 nW m$^{-2}$ sr$^{-1}$ established by Stanev \& Franceschini
(1998) measuring the optical depth at high energies due to the $\gamma
\longrightarrow \gamma$ interaction with the background infrared
photons and the lower limit of $\nu I(\nu) |_{15 \mu \rm m} = 2.4$ nW
m$^{-2}$ sr$^{-1}$ obtained by Elbaz et al. (2001) integrating the
flux of all the sources in the deep ISOCAM surveys down to the flux
limit of 0.05 mJy.  Franceschini et al. (2001), on the basis of their
evolutionary model, which takes into account counts in the mid-IR, far-IR
and sub-mm, and measurement of the far-IR background, expect that the
contribution of fainter sources would bring the total background to
$\nu I(\nu) |_{15 \mu \rm m} = 3.3$ nW m$^{-2}$ sr$^{-1}$.  This value,
which is not far from values predicted by other models (Chary \& Elbaz
2001, Xu 2000) and from values found by Altieri et al. (1999) using
cluster-lensed data, has been adopted in our analysis.

The horizontal band in Fig.~\ref{fig:alpha} represents the
$\alpha_{\rm IX}$ of the cosmic background, assuming that most of the flux
in the two spectral windows comes from sources with a similar
distribution of redshifts centred around $z=1$. Therefore, this value
should correspond to the mid-IR to X-ray index of the population which
dominates the X-ray background if the same population were
responsible for the totality of the mid-IR background. Otherwise,
fitting both backgrounds requires a combination of AGN and star
formation activity.
 
The flattening of the 2--8 keV counts in the HDF deep Chandra survey
(Brandt et al. 2001a) clearly shows that almost all the hard X-ray
background is resolved at the sensitivity of this survey. Since these
counts agree very well with the counts by Mushotzky et al. (2000),
extrapolating their result we can say that about 85\% of the 2--10 keV
background is resolved at a flux of $0.5\times10^{-15}$ erg s$^{-1}$
cm$^{-2}$ (see also Alexander et al. 2001a). We do not consider
sources with fluxes less than this value because most of them have only
upper limits on the flux and are probably starburst galaxies
(according to their low X-ray luminosity).

To evaluate the AGN contribution to the mid-IR background we divide
the sources in two groups  according to their X-ray fluxes: sources
brighter than $10^{-14}$ erg s$^{-1}$ cm$^{-2}$ and faint sources
with 2--10 keV flux in the range $0.5\times10^{-15}$---$10^{-14}$ erg s$^{-1}$
cm$^2$. In these flux ranges the sources have similar $\alpha_{\rm IX}$ 
values (see Fig.~\ref{fig:alpha}).  Using the counts of Brandt et
al. (2001a) and the results of Ueda et al. (1999) and Mushotzky et
al. (2000), sources brighter than $10^{-14}$ erg s$^{-1}$ cm$^{-2}$
contribute 40 $\pm$ 10\% of the hard X-ray background, while sources with flux
in the range $0.5\times10^{-15}$---$10^{-14}$ erg s$^{-1}$ cm$^2$
contribute 45 $\pm$ 5\% of the hard X-ray background.

We can evaluate the AGN contribution to the mid-IR background by
means of the median spectral indices of bright and faint X-ray
sources. Bright sources, most of them are in the Elais-S1 survey, have a
median $\alpha_{\rm IX}$ of 1.15, which corresponds to only 6\% of the
value required to fill the mid-IR background. Therefore, bright hard
X-ray sources contribute to the mid-IR background
$(40\pm10)\%\times6\%=(2.4\pm0.6)\%$, i.e. in a negligible way.

The median value of $\alpha_{\rm IX}$ for faint sources is 1.30, which
corresponds to 33\% of the mid-IR background. Hence, faint hard X-ray
sources contribute to the mid-IR background
$(45\pm5)\%\times33\%=(14.8\pm 1.7)\%$. Combining these results, we
conclude that sources making up $\sim$85\% of the 2--10 keV background
contribute $(17.2\pm2.3)\%$ of the mid-IR background.

\section{Summary and conclusions}
We have presented the cross-correlation between mid-IR and X-ray
observations in the Lockman Hole- and HDFN-centred regions.  ISOCAM and
XMM--Newton observed a common region of more than 200 square arcminutes
in the Lockman Hole.  A total of 24 galaxies out of 76 XMM--Newton
sources in this field show mid-IR emission.  In particular, the
percentage of hard X-ray sources with 15$\ \mu$m emission is around
60\%.  On the other hand, only around 10\% of the mid-IR sources show
X-ray emission in the different XMM--Newton bands. Deep Chandra
observations (Brandt et al. 2001a) completely cover the ISOCAM
observations of the HDF and flanking fields.  In a region of 24 square
arcminutes, 25\% of the mid-IR sources have been detected in the X-ray
for a total of 24 sources.  A comparison of the Lockman Hole, HDFN 
and Elais-S1 surveys (Alexander et al. 2001) shows that these surveys
are compatible in terms of source density taking into account their
respective detection limits.  While the HDFN survey is so
sensitive to the detection of even normal galaxies and the Elais-S1 survey
detects only very powerful and rare type-1 AGNs, the Lockman Hole survey is able
to detect a  population of galaxies whose emission is
mostly dominated by AGNs of types 1 and 2. In particular, thanks to the
increased sensitivity of XMM--Newton with respect to ROSAT,  nearly
half of the sources with mid-IR emission are new XMM--Newton sources.
Most of the sources which are optically studied are type-1
AGNs. Relying on optical colours and X-ray hardness ratio diagrams, we
conclude that about 70\% of the detected sources are type-2 AGNs.
Nevertheless, XMM--Newton observations are not  deep enough to detect
all the obscured AGNs in the sample as are the Chandra observations in the
HDF.  Only forthcoming observations of the Lockman Hole with
XMM--Newton will be able to detect the population of faint obscured
X-ray sources visible with Chandra in the HDF.  We have studied how
the mid-IR to hard X-ray index ($\alpha_{\rm IX}$) varies with 
redshift, comparing it with the expected behaviour from local
templates.  Most of the Lockman  sources detected lie in a region of
the diagram occupied by type-2 AGN local templates and several  are
optically classified as type-1 AGN. Since in general these sources do
not appear highly extincted using X-ray hardness diagnostic diagrams,
a possible explanation is that  emission from the host galaxies
contributes a fraction of their IR-optical emission greater than
that of local templates.

Finally, we have evaluated how much the integrated emission of AGN
contributes to the total extragalactic mid-IR background light using
two independent methods.  A direct estimation gives a percentage of
$(15\pm5)\%$ for the Lockman survey ($0.5<F_{15\ \mu m}<3$ mJy)
and a value of $(18\pm7)\%$ for the HDFN survey ($0.1<F_{15\ \mu
m}<0.5$ mJy), hence a contribution of $(17\pm6)\%$ in the
interval of fluxes $0.1<F_{15\ \mu \mbox{m}}<3$ mJy.  Considering
median mid-IR to X-ray spectral indices for two hard-X flux ranges, we
estimate that the population of AGNs making up $\sim$ 85\% of the 2--10
keV X-ray background contribute $(17\pm2)\%$ of the mid-IR
extragalactic background.

This fraction could be higher if there exists a population of AGNs that is
highly obscured at X-ray wavelengths. In particular, since the X-ray
background peaks at 30--40 keV while we can now observe only up to 5--10
keV, we expect that deeper X-ray observations (over a wider spectral
range) will unveil more highly extincted AGNs. 


On the basis of the presently available mid-infrared and X rays
observations, it appears that the bulk of the mid-IR extragalactic
emission comes from star formation and that the luminous galaxies seen
by ISOCAM in the deep surveys are essentially starbursts obscured by
dust. Nevertheless, this result does not exclude the possibility that
the majority of the galaxies in the Universe have both AGN and
starforming contributions.  Results obtained by Elbaz et al. (2001)
about the origin of the infrared background light based on a set of
observed correlations and by Flores et al. (1999) evaluating the
mid-IR part of the star formation are not significantly affected by
the AGN contribution to the mid-IR extragalactic light.

\begin{acknowledgements}
F.D. dedicates this work to the memory of his professor Giuliano
Giuricin, recently deceased, who introduced him to the study of AGNs.
F.D. acknowledges support from the network \emph{ISO SURVEY} set up by
the European Commission under contract ERBFNRXCT960068 or its TMR
program. H.F. was supported by a grant of the ``Academie de la
science''. We thank the anonymous referee for his carefully reading of
the manuscript, interesting comments, and suggestions which greatly
improved the paper. We are grateful to H. Aussel for providing us his flux
list before publication. We also thank M. Arnaud, I. Perez-Fournon and
F. La Franca for fruitful discussions and interesting suggestions.
\end{acknowledgements}

\end{document}